\documentstyle[oldlfont]{l-aa}
\input psfig

\newcommand{\natmine}[2]{Nat #1, #2}
\newcommand{\apjm}[2]{ApJ #1, #2}
\newcommand{\apj}[2]{ApJ #1, #2}
\newcommand{\apjsm}[2]{ApJS #1, #2}

\newcommand{\ajm}[2]{AJ #1, #2}

\newcommand{\aaa}[2]{A\&A #1, #2}

\newcommand{\aeta}[2]{A\&A #1, #2}

\newcommand{\aetas}[2]{A\&AS #1, #2}

\newcommand{\paspmine}[2]{PASP #1, #2}
\newcommand{\mn}[2]{MNRAS #1, #2}
\newcommand{\lam}{$\lambda$}
\newcommand{\about}{$\approx$}
\newcommand{\nh}{\hbox{N$_{\rm H}$}}
\newcommand{\ion}[2]{\hbox{#1\,{\sc#2}}}

\newcommand{\Halpha}{H${\alpha}$}

\newcommand{\Hbeta}{H${\beta}$}

\newcommand{\ergcm}{erg\,cm$^{-2}$\,s$^{-1}$}
\newcommand{\ergs}{erg\,s$^{-1}$}

\newcommand{\Lx}{$\rm L_{\rm X}$}

\newcommand{\Lbol}{\rm L$_{\rm bol}$}

\newcommand{\degre}{\degr}
\newcommand{\degree}{\degr}
\newcommand{\cnts}{cts\,s$^{-1}$}
\begin{document}

   \thesaurus{06         % A&A Section 6: Form. struct. and evolut. of stars
              (13.25.5;  % X-ray stars
               08.05.1;  % Stars: early,
               08.05.2;  % Stars: emission-line, Be
               08.23.1;  % Stars: white dwarfs
               08.14.1;} % Stars: neutron,
   \title{New massive X-ray binary candidates from the ROSAT Galactic Plane
Survey\thanks{Partly based on observations obtained at the
Observatoire de Haute-Provence, CNRS, France, on observations
collected at the European Southern Observatory La Silla, Chile, with
the 2.2\,m telescope of the Max-Planck-Society and on observations
acquired at the Laboratorio Nacional de Astrofisica, CNPq, Brazil}
    }
   \subtitle{I - Results from a cross-correlation with OB star catalogues}

   \author{C. Motch \inst{1,2}
       \and    F. Haberl  \inst{2}
       \and    K. Dennerl  \inst{2}
       \and    M. Pakull   \inst{1}
       \and    E. Janot-Pacheco \inst{3}
          }

   \institute{
              Observatoire Astronomique, UA 1280 CNRS, 11 rue de l'Universit\'e,
              F-67000 Strasbourg, France
              \and
              Max-Planck-Institut f\"ur extraterrestrische Physik, D-85740 
              Garching bei M\"unchen, Germany
              \and
              Instituto Astron\^omico e Geof\'isico, Universidade de S\~ao Paulo,
              Caixa Postal 9638, 01065-970 S\~ao Paulo, Brazil
              }

   \date{Accepted for publication in Astronomy \& Astrophysics Supplement Series}

   \offprints{C. Motch}

   \maketitle
 
   \begin{abstract}
	
     We report the discovery of several new OB/X-ray accreting binary candidates. 
These massive systems were found by cross-correlating in position SIMBAD OB star
catalogues with the part of the ROSAT all-sky survey located at low galactic latitudes
($|b|\, \leq 20$\degre ) and selecting the early type stars which apparently displayed
the most significant excess of X-ray emission over the `normal' stellar level. The
present search is restricted to stars earlier than B6 and X-ray luminosities $\geq$
10$^{31}$\,\ergs . Follow-up optical and X-ray observations allowed to remove
misidentified OB stars and spurious matches with interloper X-ray emitters (mostly
active coronae) leaving five very likely new massive X-ray binaries: the O7 star
LS~5039 and the Be stars BSD~24-~491,  LS~992, LS~1698 and LS~I~+61~235. This latter
source was already mentioned in an earlier paper. LS~1698 is the probable optical
counterpart of the hard X-ray transient 4U~1036-56.

These new candidates have 0.1-2.4 keV un-absorbed luminosities $\geq$ 2 10$^{33}$
\,\ergs \ indicating an accreting neutron star or black hole. On the average their soft
X-ray luminosities are comparable to those observed from hard X-ray transients in
quiescence or from persistent low luminosity Be/X-ray sources. The four Be
stars have Balmer emission slightly less intense than previously known systems
showing strong outbursts. This suggests that the relative weakness of the
circumstellar envelope may explain the low luminosities to some extent.  

Two additional X-ray binary candidates, HD~161103 and SAO~49725 require further
confirmation of their X-ray excess.  Their lower soft X-ray luminosities (1-5
10$^{32}$\,\ergs) could qualify them as Be + accreting white dwarf systems.  

Four other B stars in the Orion and Canis Major OB associations, HD 38087, HD
38023, HD 36262 and HD 53339 exhibit X-ray flux excesses in the range 2-7
10$^{31}$\,\ergs \ whose origin is unclear. Finally very soft X-ray emission was
detected from HR 2875 suggesting the presence of a non-accreting white dwarf
companion to the B5 star.  

   \end{abstract}

\keywords{X-ray: stars, Stars: early, Stars: emission-line, Be, neutron stars, white dwarfs}

%
%  14.Sep.'90: Demo-Vs.
%________________________________________________________________

\section{Introduction}

Massive X-ray binaries were among the very first X-ray sources detected and optically
identified more than 20 years ago. These systems consist of a compact object, a
magnetized neutron star (X-ray pulsar) or a black hole, in orbit around a massive OB
star. One usually distinguishes three subtypes depending on whether the X-ray source
accretes matter through Roche lobe overflow via an accretion disc (e.g.
LMC X-4), from the high velocity wind of an early type star (e.g. Vela X-1) or from a
low velocity extended envelope around a Be star (e.g. A0535+26).  Because Bondi-Hoyle
accretion (Bondi \& Hoyle 1944) is extremely sensitive to the relative velocity of the
orbiting compact object with respect to the circumstellar material, the low
velocities and high densities usually observed in Be envelopes provide rather favourable
conditions for accretion and not surprisingly, the majority of the known massive X-ray
binaries are in fact Be/X-ray systems (see Van den Heuvel \& Rappaport 1987 and Apparao
1994 for recent reviews).  

Sporadic ejection of matter often observed in single Be stars combined with large
variations of the accretion radius along the eccentric orbit explain the fact that
most of these objects appear to be highly variable or even transient sources. The
known massive X-ray binaries exhibit a large range of luminosities  (10$^{33 - 38}$
\,\ergs ; 1-10\,keV). In spite of their concentration at low galactic
latitude, their hard X-ray spectra resulting from the accretion onto a highly
magnetized young neutron star or in rare cases onto a black hole allows to detect them
very deeply into the galactic plane especially during the outburst states.  However,
large interstellar absorption often renders optical identification very difficult.

As individual objects, their scientific importance is high since, for instance, a
large fraction of our knowledge on the masses of neutron stars comes from the Doppler
analysis of these X-ray pulsars. By studying the variation of the X-ray luminosity and
absorption by the intervening medium along the orbit, i.e. by using the neutron star
as a probe of the stellar environment, several characteristics of the circumstellar
envelopes (e.g.  Waters et al. 1989; Motch et al. 1991a) or of the stellar wind (e.g. 
Haberl et al.  1989, Haberl \& White 1990) can be constrained.  

As a class of objects, their scientific importance is also large since they could
account for part, if not all, of the hard diffuse emission observed in our Galaxy
(e.g. the hard X-ray ridge emission; Warwick et al. 1985) and more generally in
starburst galaxies (e.g. Griffiths \& Padovani 1990) where they may turn on as bright
X-ray sources already 9 10$^{6}$ yr after the onset of star formation.  Finally, these
binaries may end their evolution by forming high mass binary pulsars (see e.g. 
Verbunt \& van den Heuvel, 1995 for a review).

With a 1-10\,keV sensitivity of the order of 2.5 10$^{-11}$\,\ergcm,  X-ray surveys
carried out before ROSAT had detection threshold luminosities of \Lx \ $\approx$ 3
10$^{33}$ (d/1\,kpc)$^{2}$\,\ergs . Therefore, many high mass systems may still remain
hidden in the galactic plane, especially those powered by high velocity wind accretion
and those belonging to the low end of the Be/X-ray luminosity function. Although the
low energy range of the ROSAT PSPC may not be the best suited for searching such
objects in regions of high interstellar absorption, the ROSAT all-sky survey has
nevertheless the capability to detect these systems up to distances 10 times larger
than previous experiments in directions relatively clear of interstellar absorption.

In this paper we present the result of a search for new OB/X-ray systems using the
cross-correlation in  position of all O and B stars listed in the SIMBAD database with
the source list of the ROSAT galactic plane survey (RGPS; Motch et al. 1991b). By
definition the RGPS is the part of the ROSAT all-sky survey (RASS; Voges 1992)
restricted to regions of absolute galactic latitudes below 20\degree.  The ROSAT
satellite and instrumentations are described in Tr\"umper (1983) and Pfeffermann et
al. (1986). 

This work extends the preliminary search carried out by Meurs et al. (1992) for a
subsample of early type stars.  By selecting stars which apparently exhibited a
\Lx/\Lbol \ ratio in excess of what is expected from normal stellar OB X-ray emission
we could isolate a subset of candidate objects.  Follow-up optical and pointed ROSAT
observations allowed to assess the reality of the X-ray luminosity excess and resulted
in the probable discovery of five new massive X-ray binaries with an additional two
likely candidates.  

In section 2 we present the methods and results of the initial selection using SIMBAD
entries and the ROSAT survey source list. In sections 3 to 5 we analyze our optical
and X-ray survey and pointed ROSAT observations and draw conclusions on the reality of
each new candidate massive X-ray binary. We then discuss the nature of the new systems
discovered and their relation to the already known population of massive X-ray
binaries. A preliminary report on this work was given in Motch et al. (1996c,d).

\section{Cross-correlation of the ROSAT all-sky survey source list with SIMBAD OB
stars}

\subsection{The ROSAT source list}

The source list considered here arises from the first ROSAT all-sky survey data
reduction as completed by 1991 October by the Scientific Analysis System Software
(SASS; Voges et al. 1992). This software provides for each source the position, count
rates and hardness ratios HR1 and HR2 defined as 

\begin{displaymath} {\rm HR1}\ = \
\frac{(0.40-2.40)-(0.07-0.40)}{(0.07-2.40)} 
\end{displaymath} 
\begin{displaymath} 
{\rm HR2}\ = \ \frac{(1.00-2.40)-(0.40-1.00)}{(0.40-2.40)} 
\end{displaymath} 

where (A-B) is the raw count rate in the energy range A to B in keV.  Because of
spacecraft problems no data were available for about 5\% of the sky located
between ecliptic longitudes 41\degree \ and 49\degree \ and ecliptic longitudes
221\degree \ and 229\degree . The accumulation by 2\degree \ wide strips along the
great scan circles yields variations of sensitivity perpendicular to the strip and
overlapping at high ecliptic latitudes may produce multiple detection of the same
source in adjacent strips. The lists of sources derived from each strip were then
merged into a single master list totaling about 15,000 sources at $|b|$ $\leq$
20\degree . The merging process assumed that a source detected in two or more strips
was the same when the difference in position was less than a minimum value of one
survey sky pixel (90\arcsec) or less than the combined positional errors.  The strip
oriented analysis does not allow an easy estimate of the sensitivity of detection in
a given part of the sky but allows quick detection.  The errors on ROSAT X-ray
positions have two different origins. First, the statistical uncertainty with which
the centroid of the X-ray image is positioned on the pixel grid by the Maximum
Likelihood source detection algorithm.  Second, the systematic error in the knowledge
of the satellite attitude for each photon collected in scan mode. The analysis of the
subsample of the 13 known high mass X-ray binaries (HMXBs) detected in the RGPS by
the SASS (see section 2.2 and Table \ref{knownxrbs}), led to the conclusion that the
systematic attitude error $\Delta_{att}$ to apply to survey positions was $\approx$
8\arcsec \ and that the 95\% confidence radius could be expressed as

\begin{displaymath} 
{\rm r}_{95}  \ = \ 2.5 \times \sqrt{ \Delta^{2}_{xy} \ +\Delta^{2}_{att} }
\end{displaymath} 
where $\Delta_{xy}$ is the Maximum Likelihood error.  

\subsection{The SIMBAD database}

The master list used for the cross-correlation was extracted from the
SIMBAD database in 1990 and not later updated.  SIMBAD early type stars are mostly
recognized from HD, CD, BD, HR and SAO spectral information and associated literature.
We list in Table \ref{simbadtypes} the distribution in spectral types of the SIMBAD
stars located within 20\degree \ from the galactic plane, subdwarfs excluded.  Stars
having general spectral types `OB' mostly originate from the Luminous Star (LS)
catalogues compiled by the Hamburger Sternwarte and Warner and Swasey Observatories for
both hemispheres (Hardorp et al. 1959, Stephenson \& Sanduleak 1971). Another
important group of `OB' stars without precise spectral classification arises from the
Vatican Emission line Star (VES) catalogue (McConnell \& Coyne 1983). Most of the stars
classified as `OB' are in fact earlier than B3 (Slettebak \& Stock 1957). Because of
the heterogeneousness of the various OB catalogues, it is difficult to quantify with
accuracy the completeness in magnitude of our input sample. The Luminous
Star catalogue which gathers most of the faintest early type stars is apparently
complete down to B $\approx$ 12 over the whole galactic plane.

\begin{table}
\caption{Distribution in spectral types of the OB SIMBAD stars ($|b| \leq $20\degree) }
\label{simbadtypes}
\begin{tabular}{lr}
Types  & Number \\
\hline
O  & 702 \\
B0 - B5 & 8016 \\
B6 - B9 & 20194\\
`OB'    & 7177 \\
\end{tabular}
\end{table}

We first extracted all SIMBAD entries which had an associated error circle overlapping
a 3\arcmin \ wide box centered on each ROSAT source.  This rather large search area
allows the correlation of candidate OB/X-ray sources with objects having inaccurate
coordinates such as some variable stars, or with supernova remnants.  In a second step
we selected from the main correlation list all sources having a match within 1\arcmin \
with a star of spectral type O or B, retaining for each selected OB/X-ray association
the entire list of possible candidates extracted in the 3\arcmin \ wide box.  The
restriction on angular distance allows to eliminate many spurious OB/X-ray matches
since the 90\% confidence ROSAT radius is usually less than 30-40\arcsec \ (Voges et
al.  1992, corresponding to a 95\% confidence radius in the range of 35-46\arcsec ) and
optical positions of OB stars are known to better than $\approx$ 10\arcsec .  Thirteen
sources which were thought to have a likelier identification than the proposed OB star
were removed by hand from the correlation list (see Table \ref{rejectedlist}).  This
happened for instance when a known active corona (RS CVn binary, pre-main sequence
star) or a supernova remnant was also present in the ROSAT error circle. Pre-main
sequence stars and RS CVn binaries are known to be sometimes bright soft X-ray sources
with luminosities up to 10$^{31}$\,\ergs \ (e.g.  Montmerle et al.  1983, Walter \&
Bowyer 1981). Keeping in mind that our aim was to select only promising accreting
candidates leaving doubtful identifications for a later study we decided to ignore
these cases for the moment. On similar grounds, compact OB associations (i.e.  groups
of OB stars located within few arcminutes and appearing blended at the X-ray spatial
resolution of the RASS) were not considered in this analysis because of the difficulty
in assessing an individual X-ray to bolometric luminosity ratio for such objects.  The
B2V star HD 63177 (V = 8.31) tentatively associated with RX\,J0744.9$-$5257 was also
taken off from the final correlation list since optical follow-up observations have
shown the presence of a cataclysmic variable $\approx$ 40\arcsec \ away from the
candidate B star (Motch et al.  1996a).

\begin{table*} 
\caption{List of OB/X-ray associations removed from the main correlation list on the
basis of the existence of a possible alternative identification. Count rates are taken from
the SASS analysis of the survey and the optical information listed
is entirely extracted from SIMBAD as in 1990}
\label{rejectedlist}
\begin{tabular}{lrllrllr}
\hline
ROSAT Source & count rate & \multicolumn{2}{c}{Proposed Identification} & d$_{\rm O-X}$  
&\multicolumn{2}{c}{OB match} & d$_{\rm O-X}$ \\
name        & (10$^{-3}\rm s^{-1}$) 
            & Name 
            & Sp. type
            & (\arcsec) 
            & Name 
            & Sp. type 
            & (\arcsec) 
\\
\hline
% 10 22 12.9 -58 16  2.5 
RX J1022.2$-$5816 & 99$\pm$47 &HD 90074     & G6III     &21  &  HD 302780 & B9V  & 58\\      % 574

%22 10 55.2 +63 23 44.4 
RX J2210.9+6323 & 27$\pm$14 &ADS 15712 C  &           &8   &  HD 210808 & B5   & 32\\     % 773

%21 01 38.0 +68 09 26.2 
RX J2101.6+6809 & 49$\pm$12 &HZ Cep       & Flare star&24  &  HD 200775 & B2Ve & 23 \\     % 872

%23 29 53.3 +58 33 45.6 
RX J2329.8+5833 & 49$\pm$19 &HD 221237    & A1V       &52   & ADS 16795G& B3   &54  \\     % 943
                &           &+ADS 16795H/F ?&  & & & & \\
%15 20 12.2 -38 22  4.4 
RX J1520.2$-$3822 & 150$\pm$24&CD-37 10147C/D/B &            &11& HD 136125 & B9V &46 \\     % 2571

%03 44 34.3 +32 08 49.1 
RX J0344.5+3208 & 166$\pm$25&LkHA 95      & PMS            &59&  HD 281159 & B5V &57 \\     % 2965

%03 45 37.7 +32 26 38.1 
RX J0345.6+3226 &  40$\pm$14&LkHA 329     & K5IV-Ve    & 45& HD 281157 & B5 & 22  \\     % 2967

%15 44 21.3 -41 49 23.5 
RX J1544.3$-$4149 & 116$\pm$22& HR 5846      & A0V       &19 &  HD 140285B& B  &24  \\     % 3432

%16 12 15.7 -28 25 10.3 
RX J1612.2$-$2825 & 260$\pm$45& ADS 9953 B   & F2V       & 32 &  HR 6029   & B9V &8  \\     % 3850

%05 10 22.0 +35 47 51.8 
RX J0510.3+3547 & 57$\pm$24& HP Aur$^{1)}$    & G0 (RS CVn)&2&  HD 280603 & B  & 25  \\     % 5216 (Fusion faite ds Simbad)

%05 35 59.3 -06 16 16.0 
RX J0535.9$-$0616 & 90$\pm$22& V1178 Ori    &            &6&  HD 37131  & B9V &14 \\     % 6123

%18 11 30.0 -19 25 38.6 
RX J1811.5$-$1925 & 680$\pm$70& SNR 011.2-00.3& SNR      &15  &  LS 4738   & B  &16   \\     % 7700

% Additional id=1371
RX J1158.3$-$5759 & 50$\pm$27&HD 104010 (+B) & G7+K0  &15.5  & HD 103996 & B4V &8  \\
\hline
\end{tabular}

{\small $^{1)}$ The latest version of SIMBAD merges these two
objects as a single RS CVn binary}

\end{table*}

Finally, for the sake of comparison, we keep aside the group of 13 known massive X-ray
binaries detected during the ROSAT all-sky survey at $|b|\ \leq \ 20$\degree \ (see
Table \ref{knownxrbs}). Of these, 9 were listed in the OB SIMBAD catalogues and
retrieved correctly during the above selection process while 4 were added using data
extracted from the literature (He 3 -640, Cen X-3, 1H1909+096, EXO 2030+375). The final
list of OB/ROSAT source associations contained a total of 237 sources entries split as
shown in Table \ref{xcorlist}.

\begin{table}
\caption{The final OB/RGPS correlation list}
\label{xcorlist}
\begin{tabular}{lrr}
Types  & limiting radius & Number of \\
       &                & X/OB associations \\
\hline
Known HMXBs & 1 arcmin & 13  \\
\hline
All types   & 1 arcmin & 224 \\
\hline
O-B5 + `OB' & 1 arcmin & 128 \\
O-B5 + `OB' & r$_{95}$ & 108 \\
%O-B5 + `OB' & r$_{90}$ & 98  \\
\hline
B6-B9       & 1 arcmin & 96  \\
B6-B9       & r$_{95}$ & 79  \\
\end{tabular}
\end{table}

\subsection{The \Lx/\Lbol \ \em diagnosis for OB stars}

All stars of spectral type earlier than $\approx$ B5 emit soft X-rays (0.2-4.0\,keV) with a
roughly constant X-ray to bolometric luminosity ratio independent of the luminosity class
and age (e.g. Long \& White 1980; Pallavicini et al. 1981).  Sciortino et al. (1990) show
that the mean value of \Lx/\Lbol \ for O stars is close to 10$^{-6.46}$ with about 10\% of
these early type stars having \Lx/\Lbol \ in the range of 10$^{-6}$-10$^{-5.5}$. From ROSAT
survey data Meurs et al. (1992) find a mean log\,(\Lx/\Lbol) of $-$6.8 $\pm$ 0.5 for 43 O3
to B2.5 stars with no strong difference between OB and OBe stars. Using ROSAT PSPC pointed
observations, Cassinelli et al. (1994) find that the \Lx/\Lbol \ ratio decreases sharply
with spectral types later than B1 and could reach $\approx$ 10$^{-9}$ at B3V.  

The physical origin of the X-ray emission is still a matter of debate. Several authors
assumed a picture stretched from the solar case in which a hot corona located close to
the stellar surface lies below the cooler, high speed velocity wind (see e.g. 
Cassinelli et al. 1981, Waldron 1984). Alternatively, Lucy \& White (1980) proposed
that blobs of high density formed in the expanding wind may produce shocks and
consequently X-rays.

\subsection{Computation of the X-ray to bolometric luminosity ratio}

Interstellar extinction may alter quite significantly the ratio of X-ray to optical
flux measured from these stars. Depending on the softness of the assumed intrinsic
X-ray spectrum the decrease of the 0.1-2.4\,keV PSPC count rate with increasing 
interstellar column density may be quicker or slower than that of the optical flux.  
For instance, the
overall PSPC count rate produced by a T = 10$^{6}$\,K thin thermal spectrum (Raymond
\& Smith 1977) will be steeply decreasing with column density and at \nh \ = 10$^{21}$
cm$^{-2}$ it will be $\approx$ 100 times more dimmed than any optical flux crossing
the same interstellar medium. On the opposite, the PSPC count rate resulting from a
power law energy distribution with a photon index of 1, typical for Be/X-ray systems,
will decrease less rapidly with \nh \ than optical radiation.  

In order to correct for interstellar extinction we computed the colour excess E(B-V)
from the spectral types and magnitudes listed in SIMBAD. Intrinsic B-V were taken from
Johnson (1966) and absolute magnitude and bolometric corrections are from Deutschman
et al. (1976) and Humphreys \& McElroy (1984). The corresponding column density was
then used to compute the PSPC count rate to un-absorbed flux conversion factor
assuming a 10$^{7}$\,K thin thermal spectrum (e.g.  Pallavicini et al. 1981). 
Finally, the X-ray luminosity was estimated using the distance derived from the colour
excess and spectral type together with the catalogue V or B magnitudes.  

Obviously, there exist several possible temperatures for normal OB X-ray emission. 
Chlebowski, Harnden \& Sciortino (1989) find temperatures in the range of 3 to 9
10$^{6}$\,K from Einstein data and Cassinelli et al. (1994) reach similar conclusions
from ROSAT observations.  By intentionally assuming a temperature on the hot side of
the distribution we avoid a systematic overestimation of the X-ray luminosity when
correcting for interstellar absorption.  Only in the rare case of a soft component
(e.g. 2 10$^{6}$\,K) seen at very low \nh \ do we expect overestimation of the
un-absorbed X-ray emission. If the intrinsic spectrum is actually described by a power
law distribution of photon index 1-2 as expected for young accreting neutron stars
(e.g. White et al. 1983) the effect on the un-absorbed luminosities is not large since
we overestimate them by at most a factor of 2 at low \nh \ and underestimate them by
the same factor at \nh \ = 10$^{22}$\,cm$^{-2}$.  Finally we note that with a
10$^{7}$\,K thin thermal energy distribution, the absorbed X-ray count rate to optical
flux ratio remains constant within a factor of 2 for a large range of \nh \ and that
therefore, the estimated X-ray to bolometric luminosity ratio is relatively unaffected
by errors on the intervening column density.  

Another problem arises from the unavoidable incompleteness of the data listed in SIMBAD. 
Several stars lack precise spectral types and/or magnitudes. Keeping in mind our concern to
select the most obvious accreting candidates, in order not to overestimate the X-ray to
bolometric luminosity ratio we used as default value B0 for all stars without subtypes and
luminosity class III for all stars lacking appropriate information. When the B-V colour was
not available we arbitrarily assumed B-V = 0.0.  These default assumptions imply that we may
have missed a fraction of the low \Lx/\Lbol \ candidates. In the few cases when a range of
spectral types and luminosity classes was available, we used the average value.  For early
type stars, the difference in intrinsic colours and bolometric corrections between two
consecutive spectral types or luminosity classes is always small compared to other
uncertainties.

\begin{figure} %\picplace{0.5cm}
\psfig{figure=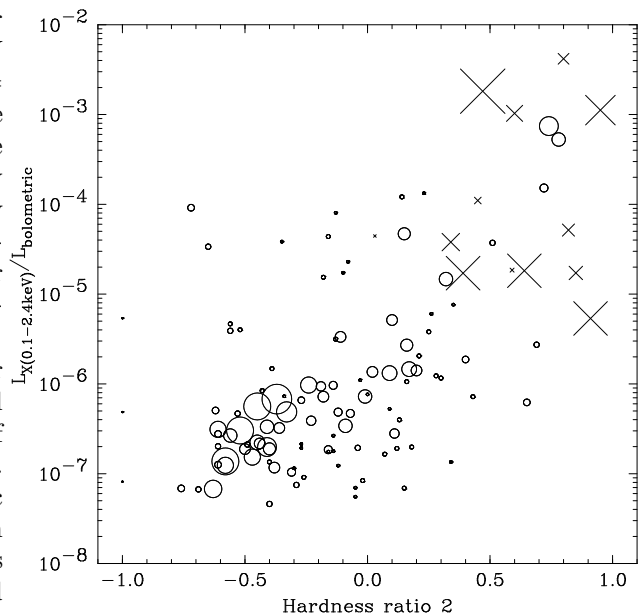,width=8.8cm,bbllx=2cm,bburx=21cm,bblly=2cm,bbury=21cm}
\caption[]{Distribution of all X-ray/OB star identifications having spectral types
earlier or equal to B5 and located within the 95\% confidence radius in the \Lx/\Lbol
\ versus HR2 diagram. Open circles represent the ROSAT OB identifications and crosses
represent the known high mass X-ray binaries detected during the ROSAT all-sky survey
and located at absolute galactic latitudes below 20\degree. The size of the symbols
is inversely proportional to the error on HR2. While most ROSAT OB star
identifications cluster in the narrow range of \Lx/\Lbol \ ratio similar to that
derived from Einstein observations,
all established X-ray binaries appear with clear X-ray excess
and much harder X-ray spectra than normal OB stars}
\label{lxxh2}
\end{figure}

We show in Fig.~\ref{lxxh2} the distribution of all X-ray/OB star identifications having
spectral types earlier than or equal to B5 and located within r$_{95}$, the 95\%
confidence radius of the ROSAT position, in the \Lx/\Lbol \ versus HR2 diagram.  The
hardness ratio HR2 is more sensitive to the intrinsic shape of the spectrum while the
softer HR1 is rather an indicator of the photoelectric absorption. Most OB stars cluster
at \Lx/\Lbol \ between 10$^{-7}$ and 10$^{-6}$ as expected from previous studies carried
out with the Einstein satellite (Pallavicini et al. 1981, Sciortino et al. 1990). 
Simulations show that indeed the slight variation in energy range from Einstein
(0.2-4.0\,keV) to ROSAT (0.1-2.4\,keV) does not significantly change the X-ray
luminosities derived from the two satellites.  Similar values of \Lx/\Lbol \ were also
derived from a subsample of ROSAT survey data (Meurs et al. 1992). We did not investigate
possible differences between OB and OBe stars in our sample. Most normal OB stars have
HR2 values comprised between $-$0.7 and +0.3 which correspond to thin thermal
temperatures in the range of 3 10$^{6}$\,K to 3 10$^{7}$\,K in agreement with those
usually reported for normal OB stars.  In contrast, the known massive X-ray binaries
detected in the galactic plane exhibit larger luminosity ratios and a much harder HR2
which probably reflects the intrinsically harder power law-like energy distribution of
accreting neutron stars and also maybe to a lower extent the often large interstellar and
intrinsic column densities. We note, however, that luminous soft X-ray components were
detected in some massive X-ray binaries accreting from the wind of the primary (e.g.  4U
1700$-$37; Haberl et al. 1994) or through Roche lobe overflow (e.g. LMC X-4; Dennerl
1991). This last source is located in a direction of low interstellar absorption and
exhibits HR1 = 0.36 $\pm$ 0.01; HR2 = $-$0.20 $\pm$ 0.01.  Because of the sometimes large
orbital phase dependent circumstellar absorption, the presence of a soft X-ray excess,
although clearly detectable from the relative strength of the hard (2-10 keV) and soft
(0.1-2 keV) un-absorbed components, does not necessarily imply a very soft value of HR1
or HR2 in the ROSAT band (e.g. 4U 1700$-$37; Haberl et al. 1994).  On the other hand,
Be/X-ray systems seem to lack a soft component (Haberl 1994).

\begin{figure}
%\picplace{0.5cm}
\psfig{figure=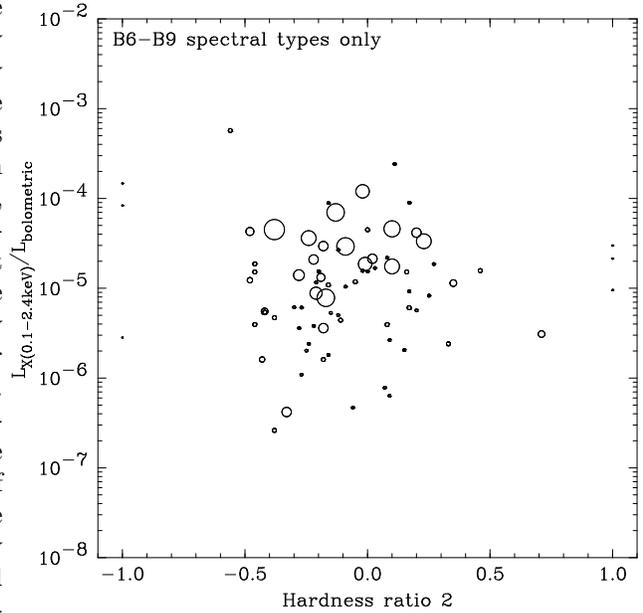,width=8.8cm,bbllx=2cm,bburx=21cm,bblly=2cm,bbury=21cm}

\caption[]{Distribution of all X-ray/OB star identifications having spectral types B6 to B9
and located within the 95\% confidence radius in the \Lx/\Lbol \ versus HR2 diagram.  The
size of the symbols is inversely proportional to the error on HR2. Late B stars do not
follow the same \Lx/\Lbol \ relation as earlier spectral types (see Fig.~\ref{lxxh2}) and
exhibit a large scatter. At the count statistics typical of survey data, their hardness
ratios HR2, however, are not essentially different from those of O-B5 stars}

\label{b9lxxh2}   % use \ref{figname} in text
\end{figure}

By contrast, it can be seen on Fig.~\ref{b9lxxh2} that B6 to B9 stars exhibit a much
larger \Lx/\Lbol \ ratio than earlier types and also a somewhat larger scatter. This
behaviour was already noticed by Meurs et al. (1992).  Obviously these late B stars do
not obey the same relation as hotter types.  The range of HR2, however, is comparable to
that observed in earlier B stars.  Einstein observations of nearby A type stars by
Schmitt et al. (1985) showed that none of these stars had any detectable X-ray emission. 
Nevertheless, the Einstein observatory detected several A type stars in the Pleiades
cluster (Micela et al. 1990) and more recently several field A stars were detected in the
ROSAT all-sky survey (Schmitt et al. 1993).  These evidences led to the common assumption
that the X-ray emission sometimes associated with late B or A stars was originating from
an optically undetectable G-M type companion. However, ROSAT HRI observations seem to
question this explanation and may point toward intrinsic X-ray emission at least in some
late B stars (Bergh\"ofer \& Schmitt 1994).  In spite of the high \Lx/\Lbol \ ratio the
actual un-absorbed X-ray luminosities all remain below 2 10$^{32}$\,\ergs \ and only four
stars exhibit X-ray luminosity in excess of 6 10$^{31}$\,\ergs . Inspection of the
optical maps of these four stars suggests the presence of likelier optical counterparts
in the ROSAT error circle.  Therefore, considering the absence of good candidates
displaying X-ray luminosities above the level at which an interpretation in terms of an
optically unseen late type companion star becomes untenable and the rather large expected
number of spurious matches resulting from the size of the entry catalogue, we decided not
to investigate the late B stars for the moment. Consequently, in the following, we shall
only consider stars earlier than B6 or those having the general `OB' type designation
totaling 15895 stars.

\subsection{Rate of spurious matches}

\begin{figure}
%\picplace{0.5cm}
\psfig{figure=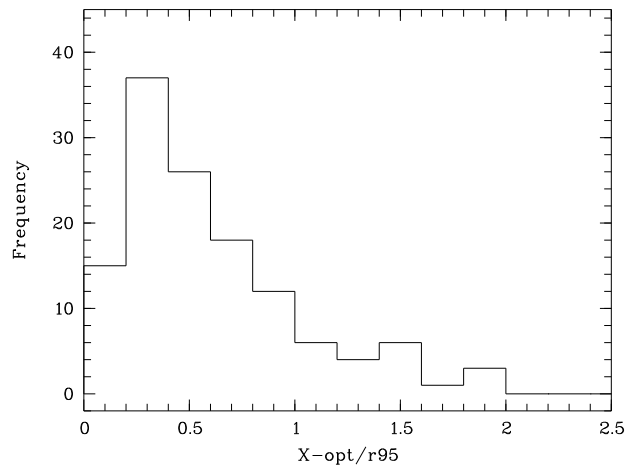,width=8.8cm,bbllx=2cm,bburx=25cm,bblly=2cm,bbury=20cm}

\caption[]{Histogram of the distances between the SASS and SIMBAD positions for the 128
stars earlier than B6 associated with a ROSAT all-sky survey source. X-ray to optical
distances are expressed in units of the 95\% confidence error radius}

\label{hidxo}   % use \ref{figname} in text
\end{figure}

We show in Fig. \ref{hidxo} the histogram of X-ray to optical distances for the 128
stars earlier than B6 associated with an X-ray source.  The shape of the distribution
shows without ambiguity that several OB stars were indeed detected during the ROSAT
survey. The average 95\% confidence radius for all OB star matches in the original
cross-correlation list is 34\farcs5. With a total of 15895 OB stars earlier than B6 and a
mean survey source density of $\approx$ one per square degree in the galactic plane
(Motch et al.  1991b), roughly independent of galactic latitude and longitude (Voges
1992), we expect 5 spurious matches among the 108 spatial coincidences within r$_{95}$
and 10 spurious matches among the 17 associations found in the range of 1 to 1.8
r$_{95}$, this latter value corresponding to the 1\arcmin \ limit.  The estimated number
of real OB/X-ray associations within and outside r$_{95}$ is 103 and 7 respectively, thus
confirming the validity of the statistics used for the computation of error radii.

Because the number of non X-ray detected OB stars is much larger than the number of OB
stars detected in the survey, we expect most spurious OB/RGPS source associations to
be with non detected early type stars and therefore to be preferentially found in the
high \Lx/\Lbol \ samples.  Consequently, the recognition of a genuine accreting source
requires additional information which may be the X-ray spectral or temporal behaviour
and/or follow-up optical searches for alternative optical counterparts, essentially 
active coronae in the galactic plane.  For the B6-B9 stars, we expect $\approx$ 6
spurious spatial coincidences among the 79 located within the 95\% confidence radius
implying that these random associations cannot explain the systematically higher
\Lx/\Lbol \ ratio exhibited by these late B stars.

\subsection{Selection of the candidate stars}

Using the computed luminosity ratio and optical / X-ray positional information we
selected three different sets of candidate OB/X-ray binaries as defined in Table
\ref{setdef}. The use of the 95\% confidence radius was guided by the fact that the
expected number of spurious matches within r$_{95}$ ($\approx$ 5) was about the same
as the number of true associations expected to be located outside this radius. 
However, when referring to the accuracy of the X-ray positions we will mention the
90\% confidence radius in order to be consistent with commonly used conventions in
X-ray astronomy.  The lower limit of \Lx/\Lbol \ = 3$\times$10$^{-6}$ is the maximum
ratio observed for normal OB stars with the Einstein satellite (Sciortino et al.
1990). As a final criterion we removed 3 stars displaying X-ray luminosities below
10$^{31}$\,\ergs \ since as already argued above, these luminosities may be radiated
by the most active late type coronae. The 3 stars (HD 37272, HD 180939 and HD 68518)
all have B5V spectral types and \Lx/\Lbol \ in the range 4-6 10$^{-6}$.  

\begin{table}
\caption{Definition of the three groups of OB/X-ray binary candidates}
\label{setdef}
\begin{tabular}{lll}
Group & \Lx / \Lbol & dist(opt-X) \\
\hline
1     & $\geq$ 10$^{-5}$ & $\leq$ r$_{95}$ \\
2     & $\geq$ 10$^{-5}$ & $>$ r$_{95}$ \\
3     & 3$\times$10$^{-6}$-10$^{-5}$ & $\leq$ r$_{95}$ \\
\end{tabular}
\end{table}

For comparison, all but one known massive X-ray binaries detected by ROSAT fall into
candidate group 1. None appears in group 2 and the only one listed in group 3 is
4U 1700$-$37 / HD 153919 which consists of a neutron star accreting in the wind of 
a hot and luminous O6.5Iaf+ star (Walborn 1973; Haberl et al. 1994). 

We then searched the literature in order to check for possible errors in the spectral
type listed in each SIMBAD object header and more generally with the aim to find
evidences for an alternative explanation to the observed X-ray emission such as a
referenced late type companion or a white dwarf.  

As a final check we analyzed interactively each remaining source with the dedicated
Extended Scientific Analysis System (EXSAS) developed at MPE (Zimmermann et al.  1992). 
Whereas SASS operated on distinct survey strips, EXSAS has the capability to use all
photons detected from a given region of the sky (1\degree$\times$1\degree \ merged fields
in our case) thus yielding improved background determination and source detection.  For
most sources, count rates, positions and hardness ratios derived from the interactive
analysis were fully consistent with those given by the SASS output. We note that some of
the positions computed by EXSAS were 10\arcsec\ or more away from the SASS determinations
but still compatible within the errors.  This slight change of X-ray position together
with the use of refined optical coordinates extracted from the Guide Star Catalogue (GSC;
Lasker et al. 1990) for 'LS' stars explains the difference between X-ray to optical
distances listed in Table \ref{sasslist} and those appearing in individual finding charts
and later Tables.  

However, in a few instances, the EXSAS process derived count rates significantly lower
than those given by SASS. In four cases (CPD $-$59 2854, LS III +58 47, HD~313343 and
LS 4287), the SASS source was hardly or even not recovered at all by EXSAS and
therefore dropped from the final list.  These discrepancies could be related to the
different data analyzed by the two processes.  Survey spectra were accumulated for
each source and light curves were systematically checked for variability and for the
presence of features characteristic of a survey artifact.

The final list of SASS/SIMBAD OB/X-ray candidates containing 24 entries 
is printed in Table \ref{sasslist}.

\begin{table*}

\caption{The list of candidate OB/X-ray binaries scheduled for optical and X-ray
follow-up observations. This table summarizes the preliminary values of spectral type,
magnitude and star distance used as input of the automatic process in order to estimate
\Lx/\Lbol \ and un-absorbed \Lx . The input information was extracted from the SIMBAD
header and results of the SASS analysis of the survey (see sections 2.1 \& 2.4). When a
range of spectral types and luminosity classes was available, we used the average value
listed here.  d$_{\rm O-X}$ is the distance between the SASS position and that of the
associated OB star as read from the SIMBAD header.  Errors listed here only arise from
the PSPC count rate and do not take into account uncertainties on the detailed spectral
type, interstellar absorption and distance. The horizontal lines divide the three groups
of candidate stars defined in Table 4}

\label{sasslist}
\begin{tabular}{llrrrcc}
Candidate      &Spectral  & B     &d$_{\rm O-X}$ & d     & \Lx /\Lbol  & \Lx  \\
Name           &Type      &       &(\arcsec)     & (kpc) &             & (erg s$^{-1}$)\\
\hline
LS I + 61 235  &  B5IIIe  & 12.15 &  7& 1.3& 5.3$\pm$0.6 10$^{-4}$ & 4.0$\pm$0.4 10$^{33}$ \\ 
BSD 24- 491    &  B0e     & 11.39 & 35& 4.1& 1.5$\pm$0.4 10$^{-5}$ & 6.8$\pm$2.0 10$^{33}$ \\ 
HD 38087       &  B5V     &  8.43 &  2& 0.5& 4.7$\pm$0.8 10$^{-5}$ & 6.9$\pm$0.7 10$^{31}$ \\ 
HD 53339       &  B3V     &  9.44 & 28& 1.1& 1.7$\pm$0.5 10$^{-5}$ & 1.3$\pm$0.4 10$^{32}$ \\
HD 59364       &  B5V     &  8.96 & 32& 1.0& 9.2$\pm$2.9 10$^{-5}$ & 1.3$\pm$0.4 10$^{32}$ \\
HR 2875        &  B5Vp    &  5.27 & 29& 0.2& 2.3$\pm$0.2 10$^{-5}$ & 3.4$\pm$0.3 10$^{31}$ \\
HD 67785       &  B2II    & 10.10 & 22& 4.0& 4.4$\pm$1.0 10$^{-5}$ & 4.9$\pm$1.1 10$^{33}$ \\
LS 992         &  B       & 12.82 & 33&11.4& 7.4$\pm$0.7 10$^{-4}$ & 3.3$\pm$0.3 10$^{35}$ \\
LS 1698        &  B       & 12.20 & 22&18.2& 1.5$\pm$0.6 10$^{-4}$ & 6.7$\pm$2.7 10$^{34}$ \\
HD 165424      &  B4II    &  9.30 & 24& 7.1& 3.8$\pm$1.7 10$^{-5}$ & 3.1$\pm$1.4 10$^{33}$ \\
LS 5039        &  B       & 12.17 &  6& 3.2& 3.7$\pm$0.9 10$^{-5}$ & 1.6$\pm$0.4 10$^{34}$ \\
LS IV -12 70   &  B       & 12.00 & 57&16.6& 1.2$\pm$0.3 10$^{-4}$ & 5.4$\pm$1.5 10$^{34}$ \\
LS III +46 11  &  B       & 12.28 & 26& 1.5& 1.5$\pm$0.2 10$^{-5}$ & 6.5$\pm$0.8 10$^{33}$ \\
\hline
BD +60 282     & B1III    & 11.02 & 54& 3.7& 1.6$\pm$0.7 10$^{-5}$ & 1.6$\pm$0.7 10$^{33}$ \\
LS I +61 298   & B        & 11.40 & 54&12.6& 1.7$\pm$0.7 10$^{-5}$ & 7.4$\pm$2.8 10$^{33}$ \\
HD 38023       & B4V      &  9.17 & 32& 0.6& 1.8$\pm$0.6 10$^{-5}$ & 4.2$\pm$1.4 10$^{31}$ \\
LS V +29 14    & B        & 11.20 & 41&11.5& 2.6$\pm$1.3 10$^{-5}$ & 1.1$\pm$0.5 10$^{34}$ \\
LS 3122        & B        & 11.40 & 47&12.6& 8.8$\pm$2.4 10$^{-5}$ & 3.9$\pm$1.1 10$^{34}$ \\
SS 73 49       & Be       & 13.50 & 57& 2.1& 2.6$\pm$0.9 10$^{-5}$ & 1.1$\pm$0.4 10$^{34}$ \\
LS 5038        & B        & 11.00 & 51&10.5& 2.9$\pm$1.1 10$^{-5}$ & 1.3$\pm$0.5 10$^{34}$ \\
LS IV -05 35   & B        & 11.90 & 28&15.8& 8.6$\pm$2.4 10$^{-5}$ & 3.8$\pm$1.1 10$^{34}$ \\
\hline
HD 36262       & B3V      &  7.52 & 17& 0.6& 5.1$\pm$1.5 10$^{-6}$ & 4.0$\pm$1.2 10$^{31}$ \\
HD 161103      & B2IVe    &  9.13 & 28& 0.9& 7.6$\pm$4.1 10$^{-6}$ & 1.8$\pm$1.0 10$^{32}$ \\
SAO 49725      & B0e      &  9.60 & 12& 2.7& 3.8$\pm$1.1 10$^{-6}$ & 1.7$\pm$0.6 10$^{33}$ \\
\end{tabular}
\end{table*}

\section{Follow-up observations}

Each entry of the final SASS/SIMBAD list of candidate OB/X-ray binaries was then
scheduled for follow-up optical observations.  The main goals of the optical
observations were to search for alternative identifications close to the X-ray
position and to acquire better spectral classification of the OB candidate.  Sources
which had passed the optical investigation and which were not falling serendipitously
in one of the ROSAT pointings were the scope of a dedicated AO-3 proposal.  

\subsection{Optical}

For the southern hemisphere, optical observations were acquired mainly at ESO with the
ESO-MPI 2.2\,m + EFOSC2 (Buzzoni et al. 1984) during a run in 1992 from April 17 till 25.
Low and medium resolution spectroscopy was obtained using respectively grism \#6 (\lam
\lam 4500-7100\,\AA ; FWHM resolution \about \ 5.4\,\AA), \#3 (\lam \lam 3500-5400\,\AA ;
FWHM resolution \about \ 3.8\,\AA), \#7 (\lam \lam 3550-4780\,\AA  ; FWHM resolution
\about \ 3.0\,\AA) and \#9 (\lam \lam 5800-7000\,\AA ; FWHM resolution \about \
3.0\,\AA).  In the imagery mode U, B, V and I frames were acquired with a pixel size of
0\farcs332 and an exposure time of 2-5\,min.  In all cases, a THX \#19 1024$^{2}$ CCD
chip was used as detector. Spectra and images were corrected for bias and flat-field
using standard MIDAS routines. Two-dimensional wavelength calibration was carried out
using spectral lines of helium and argon lamps and the observation of standard stars
allowed the flux calibration of a fraction of the spectra.  Additional spectroscopy was
obtained from 1994 February 10 till 14 with the ESO 1.5\,m telescope and the Boller \&
Chivens instrument. Finally some spectra were also obtained using the Cassegrain Boller
\& Chivens spectrograph attached at the 1.6\,m telescope of the Brazilian Laboratorio
Nacional de Astrofisica (LNA; Brazopolis, Brazil).  The detector used was a
770$\times$1152 GEC CCD and the FWHM resolution was $\approx$ 9\,\AA \ in the range \lam
\lam 3800-8200\,\AA .  

Northern fields were investigated during several observing runs performed at the
Observatoire de Haute-Provence, CNRS, France, between 1990 November and 1995 January. 
This observing programme was part of a general project aiming at the optical
identification and follow-up observations of area and X-ray selected sources extracted
from the ROSAT Galactic Plane Survey. All OHP spectroscopic observations were obtained
with the CARELEC spectrograph (Lemaitre et al.  1990) attached at the 1.93\,m telescope. 
Low resolution spectroscopy ($\lambda \lambda$ 3500-7500\,\AA ; FWHM resolution $\approx$
14\,\AA ) and blue and red medium resolution spectroscopy ($\lambda \lambda$
3800-4300\,\AA ; FWHM resolution $\approx$ 1.8\,\AA , $\lambda \lambda$ 6300-6700\,\AA ;
FWHM resolution $\approx$ 1.7\,\AA) were acquired using a 260\,\AA /mm grating and a
33\,\AA /mm grating respectively. Spectra were calibrated in wavelength using arcs of
iron, neon and helium lamps.  Observations of standard stars allowed to formally
calibrate in flux all spectra.  However, bad meteorological conditions which prevailed in
some cases and the narrow slit used for some bright stars may bias the mean flux level on
occasions and therefore, the spectrophotometric flux scale should be considered with some
caution.  CCD images were collected using the standard camera at the 1.2\,m telescope. 
Most of the time the Johnson B,V,and I filters were used.  Depending on the CCD mounted,
the field of view was either 7\farcm1 $\times$ 4\farcm4 (RCA) or  6\farcm5 $\times$
6\farcm5 (TK512).

Spectral types of OB stars were derived comparing with spectral atlases in Jaschek \&
Jaschek (1987) and Walborn \& Fitzpatrick (1990) and equivalent widths listed in
Didelon (1982).  For the later spectral types we used the atlas by Turnshek et al. 
(1985).  Interstellar absorption was usually estimated from the $\lambda$ 4430 and
$\lambda$ 6284 interstellar bands corrected for atmospheric effects and following the
relations given in Krelowski et al. (1987) and Bromage \& Nandy (1973).

\subsection{X-ray}

We list in Tables \ref{xpos} and \ref{xdata} the main X-ray characteristics of the
sources derived both from survey and pointed PSPC data. Sources appear in each group
ordered by increasing right ascension. All pointed X-ray data presented below were
analyzed using the EXSAS package. Spectra were accumulated for each source and the
light curves were checked for variability. We systematically searched all major
catalogues produced by past X-ray instrumentation for a possible previous detection of
the ROSAT source.

\tabcolsep=4pt
\begin{table*}

  \caption{Positions and errors resulting from the EXSAS analysis of survey and pointed
PSPC observations of the OB/X-ray candidates selected for follow-up observations.  We
also list the difference between the X-ray pointed and optical positions (p-o) and
difference between pointed and survey positions (p-s). The optical positions used here
are the most accurate available for each OB target and not necessarily those used in
Table 4. Column (o.a.) lists the off axis angle for the pointed observations. Source
names are computed after the survey EXSAS positions. Horizontal lines divide the three
groups of candidates defined in Table 4}

    \label{xpos}
\begin{tabular}{lrrrccccrcrl}
%
%\multicolumn{1}{l}{Candidate} & \multicolumn{1}{l}{ROSAT Source} &
%\multicolumn{1}{c}{p-o}&\multicolumn{1}{c}{p-s}&
%\multicolumn{6}{c}{Source position (2000.0)} &
%\multicolumn{1}{l}{o.a.} & \multicolumn{1}{l}{Pointing }\\
%
%\multicolumn{1}{l}{Name} & \multicolumn{1}{l}{Name} & 
%\multicolumn{1}{c}{(\arcsec)}&\multicolumn{1}{c}{(\arcsec)}&
%\multicolumn{2}{c}{Survey}&\multicolumn{1}{c}{r$_{90}$}&
%\multicolumn{2}{c}{Pointed observation}&\multicolumn{1}{c}{r$_{90}$}&
%\multicolumn{1}{c}{(\arcmin)} & \multicolumn{1}{l}{Identification}\\
%
% & & & & \multicolumn{1}{c}{$\delta$ (\degree \ \arcmin \ \arcsec)} &
%\multicolumn{1}{c}{$\delta$ (\degree \ \arcmin \ \arcsec)} & &
%\multicolumn{1}{c}{$\delta$ (\degree \ \arcmin \ \arcsec)} &
%\multicolumn{1}{c}{$\delta$ (\degree \ \arcmin \ \arcsec)} & & & \\
%
\hline
 & & & & \multicolumn{6}{c}{Source position (2000.0)} \\
\multicolumn{1}{l}{Candidate} & \multicolumn{1}{l}{ROSAT Source} &
\multicolumn{1}{c}{p-o}&\multicolumn{1}{c}{p-s}&
\multicolumn{2}{c}{Survey} &\multicolumn{1}{c}{r$_{90}$}&
\multicolumn{2}{c}{Pointed observations} &\multicolumn{1}{c}{r$_{90}$}&
\multicolumn{1}{l}{o.a.} & \multicolumn{1}{l}{Pointing }\\
\multicolumn{1}{l}{Name} & \multicolumn{1}{l}{Name} & 
\multicolumn{1}{c}{(\arcsec)}&\multicolumn{1}{c}{(\arcsec)}&
\multicolumn{1}{c}{$\alpha$ (h m s)} & 
\multicolumn{1}{c}{$\delta$ (\degree \ \arcmin \ \arcsec)} & 
\multicolumn{1}{c}{(\arcsec)} &
\multicolumn{1}{c}{$\alpha$ (h m s)} & 
\multicolumn{1}{c}{$\delta$ (\degree \ \arcmin \ \arcsec)} &
\multicolumn{1}{c}{(\arcsec)}& 
\multicolumn{1}{c}{(\arcmin)} & \multicolumn{1}{l}{Identification}\\
\hline     
%                                p-o p-s
LS I +61 235 & RX J0146.9+6121  & 5 & 9 & 01 46 59.5  & +61 21 22    & 24 & 01 47 00.8  & +61 21 22    & 17 & 24  & UK 400332 \\
BSD 24- 491  & RX J0440.9+4431  & 6 & 38&04 40 57.5  & +44 31 22    & 36 & 04 40 59.9  & +44 31 51    & 17 & 0.8 & WG 400397 \\
HD 38087     & RX J0542.9$-$0218&18 & 19&05 42 59.8  & $-$02 18 46  & 23 & 05 43 00.2  & $-$02 18 28  & 19 & 35  & WG 900189/386 \\
HD 53339     & RX J0704.2$-$1123&   &   &07 04 14.1  & $-$11 23 47  & 50 &             &              &    &     &         \\
HD 59364     & RX J0728.6$-$2629&   &   &07 28 38.2  & $-$26 29 12  & 43 &             &              &    &     &         \\
HR 2875      & RX J0729.0$-$3848&   &   &07 29 05.2  & $-$38 48 15  & 20 &             &              &    &     &         \\
HD 67785     & RX J0807.2$-$5053&   &   &08 07 12.8  & $-$50 53 12  & 22 &             &              &    &     &         \\
LS 992       & RX J0812.4$-$3114&   &   &08 12 28.4  & $-$31 14 51  & 21 &             &              &    &  30 & WG 400315 \\      
LS 1698      & RX J1037.5$-$5647& 3 & 16&10 37 33.7  & $-$56 47 49  & 25 & 10 37 35.2  & $-$56 47 59  & 21 & 0.2 & WG 400394 \\
HD 165424    & RX J1806.8$-$2606&40 & 55&18 06 53.9  & $-$26 06 25  & 33 & 18 06 57.5  & $-$26 06 00  & 17 & 1.2 & WG 400396 \\
LS 5039      & RX J1826.2$-$1450&13 & 13&18 26 14.9  & $-$14 50 29  & 22 & 18 26 14.8  & $-$14 50 42  & 35 & 41  & US 400285 \\
LS IV -12 70 & RX J1830.7$-$1232&96 & 73&18 30 45.8  & $-$12 32 28  & 67 & 18 30 50.8  & $-$12 32 29  & 24 & 31  & WG 500040 \\
LS III +46 11& RX J2035.2+4651  & 6 &  1&20 35 12.9  &   +46 51 16  & 19 & 20 35 13.0  & +46 51 17    & 18 & 0.4 & WG 400393 \\
\hline
BD +60 282   & RX J0136.7+6125  &   &   &01 36 43.2  &   +61 25 59  & 29 &             &              &    &     &           \\
LS I +61 298 & RX J0234.4+6147  &19 & 40&02 34 29.1  &   +61 47 49  & 56 & 02 34 34.0  & +61 47 29    & 30 & 24  & WG 201263 \\
HD 38023     & RX J0542.3$-$0807&18 & 22&05 42 19.1  & $-$08 07 45  & 26 & 05 42 20.0  & $-$08 08 02  & 18 & 16  & US 200901 \\
LS V +29 14  & RX J0553.0+2939  &   &   &05 53 02.2  &   +29 39 11  & 24 &             &              &    &     &           \\
LS 3122      & RX J1335.5$-$6211&   &   &13 35 33.6  & $-$62 11 04  & 22 &             &              &    &     &           \\
SS 73 49     & RX J1559.2$-$4157&   &   &15 59 14.2  & $-$41 57 57  & 44 &             &              &    &     &           \\
LS 5038      & RX J1826.1$-$1321&93 & 59&18 26 11.7  & $-$13 21 57  & 35 & 18 26 14.1  & $-$13 21 09  & 25 &  14 & US 400283 \\
LS IV -05 35 & RX J1900.7$-$0503&   &   &19 00 45.7  & $-$05 03 50  & 22 &             &              &    &     &           \\
\hline
HD 36262     & RX J0531.0+1206  &   &   &05 31 02.3  &   +12 06 03  & 23 &             &              &    &     &           \\
HD 161103    & RX J1744.7$-$2713& 5 & 21&17 44 44.5  & $-$27 13 30  & 28 & 17 44 45.4  & $-$27 13 47  & 18 & 0.6 & WG 400400 \\
SAO 49725    & RX J2030.5+4751  & 5 & 13&20 30 31.1  &   +47 51 58  & 35 & 20 30 30.6  &   +47 51 46  & 18 & 0.4 & WG 400393 \\
\end{tabular}
\end{table*}
\tabcolsep=6pt

\tabcolsep=4pt
\begin{table*}

\caption{X-ray count rates and hardness ratios derived from survey and pointed
observations for the 24 sources tentatively associated with OB/X-ray candidates and
selected for follow-up observations.  All data listed here were obtained using EXSAS with
the exception of survey hardness ratios which are those listed in the SASS automatic
analysis.  For pointed observations, the maximum likelihood (ML) values are the largest
of the broad (b) and hard (h) energy bands.  For HD38087 count rates and exposure times
are the mean and sum of pointings WG 900189 and WG 900386 whereas the ML and HRs
correspond to WG900189 only. The horizontal lines divide the three groups of candidates
defined in Table 4} 

\label{xdata} 

\begin{tabular}{lrrrccrrrcc}
\hline
\multicolumn{1}{l}{Candidate} & \multicolumn{5}{c}{Survey} & \multicolumn{5}{c}{Pointed observations} \\
%                   SURVEY                                           &                   POINTING              \\
\multicolumn{1}{l}{Name}      & count rate    & ML   & exp & HR1            & HR2        & count rate  & ML  & exp &  HR 1    & HR2      \\
            & (10$^{-3}\rm s^{-1}$)       &      & (s) &            &            &  (10$^{-3}\rm s^{-1}$)    &     & (s) &           &         \\
\hline
LS I +61 235&146$\pm$18& 119  &539  &+0.89$\pm$0.10  &+0.78$\pm$0.14  & 259$\pm$18 & 4531(b)& 10749 &+0.99$\pm$0.01 &+0.70$\pm$0.02  \\ 
BSD 24- 491 &27$\pm$9& 13   &492  &+0.75$\pm$0.22  &$-$0.18$\pm$0.52& 68$\pm$6  & 652(h) & 2048  &+0.95$\pm$0.05 &+0.55$\pm$0.07  \\
HD  38087   &74$\pm$13& 73   &504  &+0.82$\pm$0.17  &+0.15$\pm$0.16  & 66$\pm$7  & 446(b) & 47009 &+1.00$\pm$0.06 &+0.09$\pm$0.04  \\
HD 53339    &23$\pm$9&  8   &474  &+0.77$\pm$0.20  &$-$0.10$\pm$0.70&                  &     &       &               &                  \\
HD 59364    &35$\pm$11& 11   &431  &$-$0.15$\pm$0.69&$-$0.72$\pm$0.34&                  &     &       &               &                 \\
HR 2875     &663$\pm$51& 402  &277  &$-$0.98$\pm$0.02&$-$0.08$\pm$0.70&                  &     &       &               &                  \\
HD 67785    &45$\pm$9& 40   &763  &+0.13$\pm$0.52  &$-$0.16$\pm$0.53&                  &     &       &               &                  \\
LS 992      &285$\pm$27& 311  &449  &+0.98$\pm$0.02  &+0.74$\pm$0.09  &  $\leq$3     &     &  19520& ---              &  ---                \\
LS 1698     &66$\pm$20& 24   &207  &+0.74$\pm$0.23  &+0.72$\pm$0.25  & 3.3$\pm$1.7& 9.1(h) & 1198  &    ---        &    ---           \\
HD 165424   &33$\pm$15&  7   &265  &+0.76$\pm$0.30  &$-$0.35$\pm$0.62& 41$\pm$5  & 251(b) & 2016  &+1.00$\pm$0.05 &+0.16$\pm$0.12  \\
LS 5039     &52$\pm$15& 27   &331  &+0.92$\pm$0.08  &+0.51$\pm$0.38  &43$\pm$6  & 57(h)  & 7714  &+0.86$\pm$0.29 &+0.78$\pm$0.22  \\
LS IV -12 70&17$\pm$10&  4   &342  &+0.60$\pm$0.33  &+0.14$\pm$0.52 &36$\pm$2.5 &  98(h) & 11896 &+1.00$\pm$0.11 &+0.31$\pm$0.08  \\
LS III +46 11&59$\pm$10& 60  &872  &   ---          &+0.32$\pm$0.14  &46$\pm$6   & 196(h) &  1257 &+1.00$\pm$0.05 &+0.46$\pm$0.12  \\
\hline
BD +60 282   &17$\pm$7& 12  &540  &   ---          &$-$0.10$\pm$0.70&                  &     &       &               &                  \\
LS I +61 298 &11$\pm$5& 7   &631  &+0.41$\pm$0.45  &+0.16$\pm$0.69  &13$\pm$2   &33(h)   &8841   & ---           &+0.30$\pm$0.18    \\
HD 38023     &27$\pm$9& 18  &474  &+0.54$\pm$0.38  &$-$0.34$\pm$0.46&14$\pm$1   & 230(b) & 15048 &+0.95$\pm$0.05 &$-$0.11$\pm$0.07  \\
LS V +29 14  &26$\pm$9& 18  &435  &+0.29$\pm$0.49  &$-$0.14$\pm$0.52&                  &     &       &               &                  \\
LS 3122      &64$\pm$19& 23  &265  &+0.60$\pm$0.33  &$-$0.26$\pm$0.51&                  &     &       &               &                  \\
SS 73 49     &46$\pm$20& 8   &177  &+0.54$\pm$0.38  &+0.01$\pm$0.71  &                  &     &       &               &                  \\
LS 5038      &17$\pm$9& 7   &323  &+0.69$\pm$0.37  &+0.50$\pm$0.53  & 1.8$\pm$0.5& 12(h)  &  8747 &   ---         &  ---             \\
LS IV -05 35 &41$\pm$12& 24  &373  &+0.32$\pm$0.48  &  ---           &                  &     &       &               &                  \\
\hline
HD 36262     &49$\pm$12& 36  &467  &+0.69$\pm$0.37  &+0.26$\pm$0.66  &                  &     &       &               &                  \\
HD 161103    &31$\pm$13& 9   &313  &+0.28$\pm$0.65  &+0.35$\pm$0.62  & 12.7$\pm$2.6& 65(h)  & 2000  &+0.94$\pm$0.17 &+0.47$\pm$0.21  \\
SAO 49725    &8$\pm$4& 7   &920  &+0.82$\pm$0.23  &+0.25$\pm$0.50  & 19.5$\pm$4.2 & 62(h)  & 1286  &+0.91$\pm$0.21 &+0.35$\pm$0.23  \\
\end{tabular}
\end{table*}

\tabcolsep=6pt

\section{Criteria for the identification of OB/X-ray candidates}

The goal of the present work is merely to extract from the ROSAT Galactic Plane Survey
the most obvious OB/X-ray candidates. As stated in section 2.2, we decided to ignore
for the moment all associations of a survey source with groups of OB stars and all
cases where there was a possible alternative candidate to the OB identification. We
applied the same strategy during the follow-up observations, dropping the candidate as
soon as another possible optical counterpart was found.  

Active coronae constitute up to 85\% of the ROSAT survey sources at $b$ = 0\degree \
(Motch et al. 1996b) and over 50\% in the overall RGPS region (Motch et al. 1991b). 
Although no direct conspicuous signature of coronal activity exists in the optical
spectra of late type stars, it is known that chromospheric and coronal activities
tightly correlate in the Sun and F-M stars in general. In particular, Einstein
observations have shown that the luminosity of the chromospheric Balmer and
\ion{Ca}{II} \ H\&K emission lines is well correlated with the soft X-ray luminosity in
Me stars (e.g.  Fleming et al. 1988) and in F-G stars (e.g.  Maggio et al. 1987).
Using medium resolution blue optical spectra of $\approx$ 100 late type stars
identified with RGPS sources with measured \ion{Ca}{II}\ re-emission, Guillout (1996)
find a very good correlation between \ion{Ca}{II}\ H\&K and soft X-ray emission with
L$_{\rm CaII} \ \sim$ L$_{\rm X(0.1-2.4)keV}^{1.05\pm0.20}$.  Because the slope in
luminosity is close to 1, there also exists a useful relation between chromospheric
fluxes and PSPC count rates $r$; F$_{\rm CaII} \ \sim r^{0.74\pm0.14}$ with a 0.1 PSPC
\cnts \ corresponding to F$_{\rm CaII (H+K)}\ \approx \ $2 10$^{-13}$ \ergcm .  These
relations hold over 2 decades in flux and over 3 decades in luminosity. A rms scatter
of about a factor 2 is present in flux and luminosity around the mean relation with a
maximum deviation of a factor 10 above or below. The intrinsic long (solar cycles) and
short (flares) time scales of variability and the non-simultaneity of the optical
follow-up / X-ray survey observations probably account for the rather large scatter. 
We checked that the measured chromospheric fluxes and PSPC count rates of all late
type stars proposed as alternative identification to the early type star were
compatible with the relation of Guillout (1996).  Because of the sometimes inaccurate
spectrophotometry we re-calibrated our blue medium resolution spectra using either
magnitudes extracted from the literature or derived from our CCD images.  

In the absence of alternative candidates, we used the X-ray luminosity as additional
criterion.  Although active coronae with X-ray luminosities above 10$^{31}$ \,\ergs \
are usually not observed in late type stars (see e.g. Rosner et al. 1985) some RS CVns
(Ottmann \& Schmitt 1992) and extreme T Tauri stars (e.g.  Montmerle et al.  1983) do
exhibit on occasions soft X-ray luminosities above 10$^{31}$\,\ergs .  Accordingly,
candidate stars with excess \Lx \ in the range of 10$^{31-32}$\,\ergs \ were indicated
as deserving further attention but are not considered as very good candidates.  

The last criterion considered is the hardness of the source. Normal OB stars usually
exhibit HR2s $\leq$ 0.3 whereas most HMXBs known prior to the launch of ROSAT have hard
HR2s $\geq$ 0.4. However, the separation power of this X-ray spectral criterion is
decreased by the presence of a soft excess in a small fraction of these OB/X-ray
systems (e.g. LMC X-4).  Furthermore, some very active coronae may also display hard
spectra (e.g. Schmitt et al.  1990) and a fraction of T Tauri stars exhibit hardness
ratios HR2 larger than 0.4 (e.g.  Neuh\"auser et al. 1995). We did not consider the
presence or absence of Balmer emission lines as a criterion since some early type
companions of known massive sources lack these spectral features.

Finally, in spite of the large galactic absorption expected in the directions occupied
by OB stars, a few bright extragalactic emitters may still shine in X-ray through the
galactic fog. These sources will also display a hard HR2 as a result of interstellar
absorption and will be optically faint (V $\la$ 18). The detection of characteristic
features in their optical spectra may be beyond the capability of the instrumentation
used for this project. Motch et al.  (1996b) estimate that the spatial density of
extragalactic emitters in the RGPS Cygnus test region is 2.7 10$^{-2}$ extragalactic
sources per square degree brighter than 0.02 \cnts \ for $b$ in the range of
$-$5\degree\ to +5\degree. If these densities are typical of the average OB star
directions, we expect $\approx$ 0.1 spurious coincidence within a typical 95\%
confidence radius between an extragalactic emitter and an OB star.  Therefore, we
cannot completely rule out the possibility that at least one of the investigated
candidate sources is actually extragalactic.

\section{Results from individual source observations}

\subsection{Accreting candidates}

\subsubsection{VES 625 = LS I +61 235 (Group 1)}

This Be/X-ray system was discovered by Motch et al. (1991b) as a result of a preliminary
investigation of the ROSAT Galactic Plane Survey. It actually stands out from Table
\ref{sasslist} as exhibiting one of the largest \Lx/\Lbol .  Optical observations
revealed only moderate Balmer emission (EW (\Halpha ) = $-$7,$-$10\,\AA ) with a distinct
absorption core on occasions (Coe et al. 1993). Follow-up ROSAT pointed PSPC observations
by Hellier (1994) showed that the source was pulsating with a 1412\,s period thus being
the slowest Be/neutron star known. Together with the nearby unusual ultrasoft 8.7\,s
pulsar (Israel et al. 1994), RX\,J0146.9+6121 = LS I +61 235 contributes to the Uhuru
source 4U 0142+61.  The energy distribution of RX\,J0146.9+6121 is hard with a power law
energy index of $-$0.6 $\pm$ 0.7 (Hellier 1994) in agreement with the very positive HR1
and HR2 measured during the ROSAT all-sky survey.  

\begin{figure}
\psfig{figure=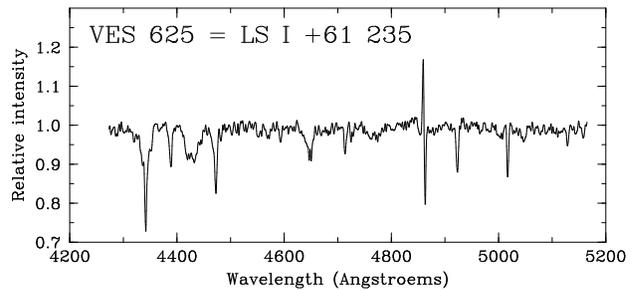,width=8.8cm,bbllx=1.5cm,bburx=23cm,bblly=2.5cm,bbury=13cm}
\caption[]{Rectified blue medium resolution spectrum of VES 625 = LS I +61 235
 obtained with the OHP 1.93\,m and CARELEC spectrograph on 1995 January 30. Exposure
time was 10\,min}
\label{spblueVES625}
\end{figure}

Medium resolution blue spectroscopy obtained at OHP on 1992 October 23 and 1995 January
30 (see Fig.~\ref{spblueVES625}) indicates a spectral type significantly earlier than the
B5III type proposed by Slettebak (1985). The relative strength of the \ion{Si}{III}
$\lambda$ 4552 / \ion{He}{I} $\lambda$ 4387 suggests a giant luminosity class.  For this
class, the \ion{C}{III} + \ion{O}{II} blend at $\lambda$ 4650 rather indicates an early B1
type.  Therefore, although our spectral range may not be ideal for spectral
classification, we believe that the star is hotter than previously thought and may be
classified as a B1IIIe type star. The revised distance of LS I +61 235 (d $\approx$
2.9\,kpc) is more consistent with the distance of the open cluster in which it is located
(d = 2.5\,kpc; Tapia et al. 1991) than that derived from a B5III type (d $\approx$
1.3\,kpc).

\begin{figure}
\psfig{figure=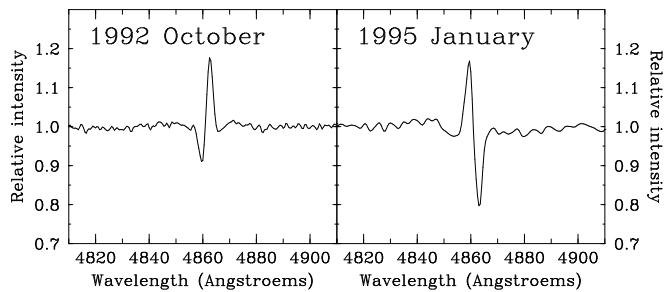,width=8.8cm,bbllx=1.5cm,bburx=23cm,bblly=2.5cm,bbury=13cm}
\caption[]{Rectified blue medium resolution spectrum of VES 625 = LS I +61 235
 obtained with the OHP 1.93\,m and CARELEC spectrograph on 1992 October 23 and
1995 January 30 showing the large V/R changes having occurred in the 2 years interval}
\label{sphbetaVES625}
\end{figure}

\begin{figure}
\psfig{figure=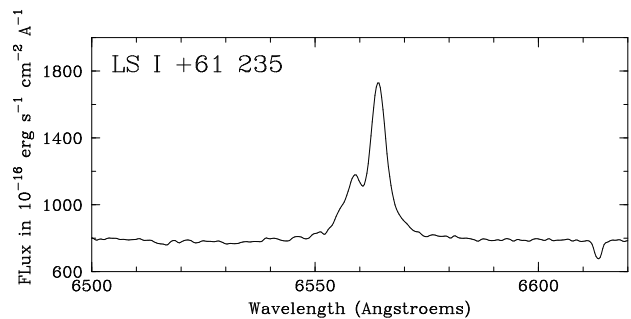,width=8.8cm,bbllx=1.5cm,bburx=23cm,bblly=2.5cm,bbury=13cm}
\caption[]{Red medium resolution spectrum of VES 625 = LS I +61 235
 obtained with the OHP 1.93\,m and CARELEC spectrograph on 1992 December 15. Exposure
time was 30\,min}
\label{spredVES625}
\end{figure}

The Balmer emission lines exhibit dramatic V/R changes on time scales of two years or
less as shown for the \Hbeta \ line in Fig.~\ref{sphbetaVES625}. Our red \Halpha \
spectrum obtained on 1992 December 15 (see Fig.~\ref{spredVES625}) is mirror-symmetric
of that plotted in Coe et al (1993) and obtained on 1991 August 28, in the sense that
the `violet' component is much stronger than the `red' one. The maximum time for the
\Halpha \ V/R reversal is of the order of 1.5 yr.

\subsubsection{BSD 24- 491 (Group 1)}

\begin{figure}
\psfig{figure=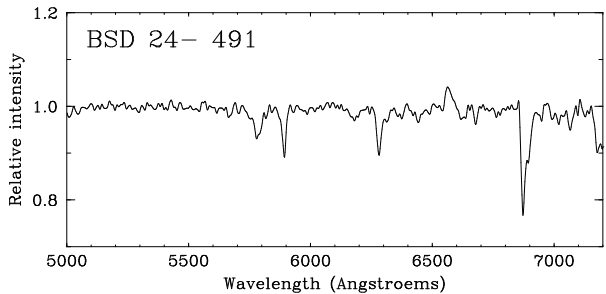,width=8.8cm,bbllx=1.5cm,bburx=23cm,bblly=2.5cm,bbury=13cm}
\caption[]{Low resolution spectrum of BSD 24- 491 obtained on 1991 November 16 UT at OHP.
The FWHM resolution is 15\,\AA . Relatively weak \Halpha \ emission is clearly detected}
\label{spredBSD24}
\end{figure}

Visual inspection of the POSS O and E plates and B and I band CCD images failed to reveal
any plausible alternative active corona counterpart within the error circles of the ROSAT
survey and pointing positions. The star BSD 24- 491 = LS V +44 17 = VES 826 is referenced
as a B0 star with indication for \Halpha \ emission (Seyfert \& Popper 1941; Coyne \&
McConnell 1983).  Low resolution spectrum obtained on 1991 November 16 (see
Fig.~\ref{spredBSD24}) shows clear evidence for \Halpha \ emission with an EW = $-$2.0
$\pm$ 0.2\,\AA .  

\begin{figure}
\psfig{figure=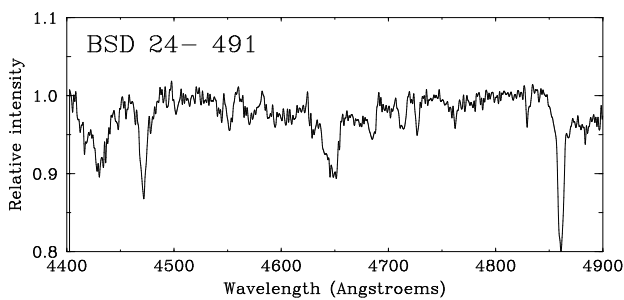,width=8.8cm,bbllx=1.5cm,bburx=23cm,bblly=2.5cm,bbury=13cm}
\caption[]{Rectified blue medium resolution spectrum of BSD 24- 491 obtained
on 1991 November 27 at OHP}
\label{spblueBSD24}
\end{figure}

A blue high resolution spectrum obtained on 1991 November 27 confirms the B0 type
listed in SIMBAD for this star (see Fig.~\ref{spblueBSD24}). There are no good
luminosity sensitive lines in our wavelength range. However, the EW of \ion{He}{I} $\lambda$
4471 rather suggests a dwarf or giant star (Didelon 1982). The photometry reported by
Bigay (1963); V = 10.78, B-V = 0.61, U-B = $-$0.36 indicates a rather high interstellar
absorption E(B-V) = 0.90 in agreement with the depth of the  $\lambda$ 4430 and
$\lambda$ 6284 interstellar bands.  An additional medium resolution spectrum obtained
at OHP in 1995 January reveals an increase of the Balmer emission compared to our
observation in 1991 with strong double peaked \Hbeta \ and iron/\ion{He}{I} lines in emission.

\begin{figure}
\psfig{figure=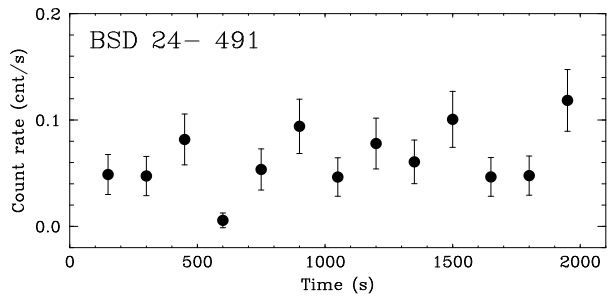,width=8.8cm,bbllx=1.5cm,bburx=23cm,bblly=2.5cm,bbury=13cm}
\caption[]{The pointed observation 0.5-2.0\,keV X-ray light curve of BSD 24- 491. Integration
time is 150 s}
\label{lcBSD24}
\end{figure}

The source was detected again during a dedicated pointed PSPC observation with a count rate
$\approx$ 2.5 times higher than during survey observations. Our pointed observation
indicates that the source is hard (HR2 = 0.55 $\pm$ 0.07) and maybe variable (see
Fig.~\ref{lcBSD24}). The position of the X-ray source derived from the pointed observation
is only 6\arcsec \ away from the GSC position of
the Be star. From all these evidences we conclude that BSD 24- 491 is very likely to be a
new Be/X-ray system.

\subsubsection{LS 992 (Group 1)}

\begin{figure}
\psfig{figure=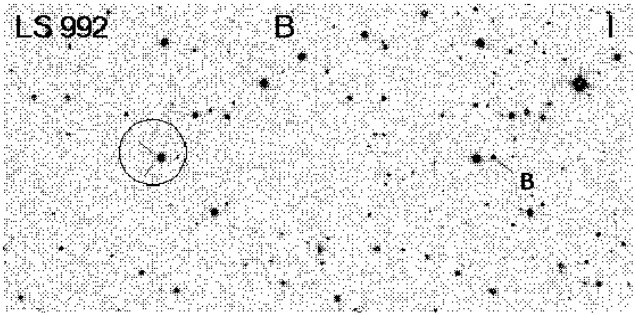,width=8.8cm,bbllx=3.5cm,bburx=18cm,bblly=10.5cm,bbury=18cm,clip=true}
\caption[]{B and I CCD images of the field of LS 992 obtained on 1992 April 18 with EFOSC2 and
the ESO-MPI 2.2\,m telescope. Both images have 2\,min exposure time.
North is at top and East to the left. Each frame is $2\farcm83$$\times$$2\farcm83$ wide. On the B image we
 plot the ROSAT survey
90\% confidence error circle and show the position of LS 992. On the I band image we show the
position of the alternative optical candidate ``B" investigated spectroscopically
and apparently unrelated to the X-ray source}
\label{imageLS992}
\end{figure}

\begin{figure}
\psfig{figure=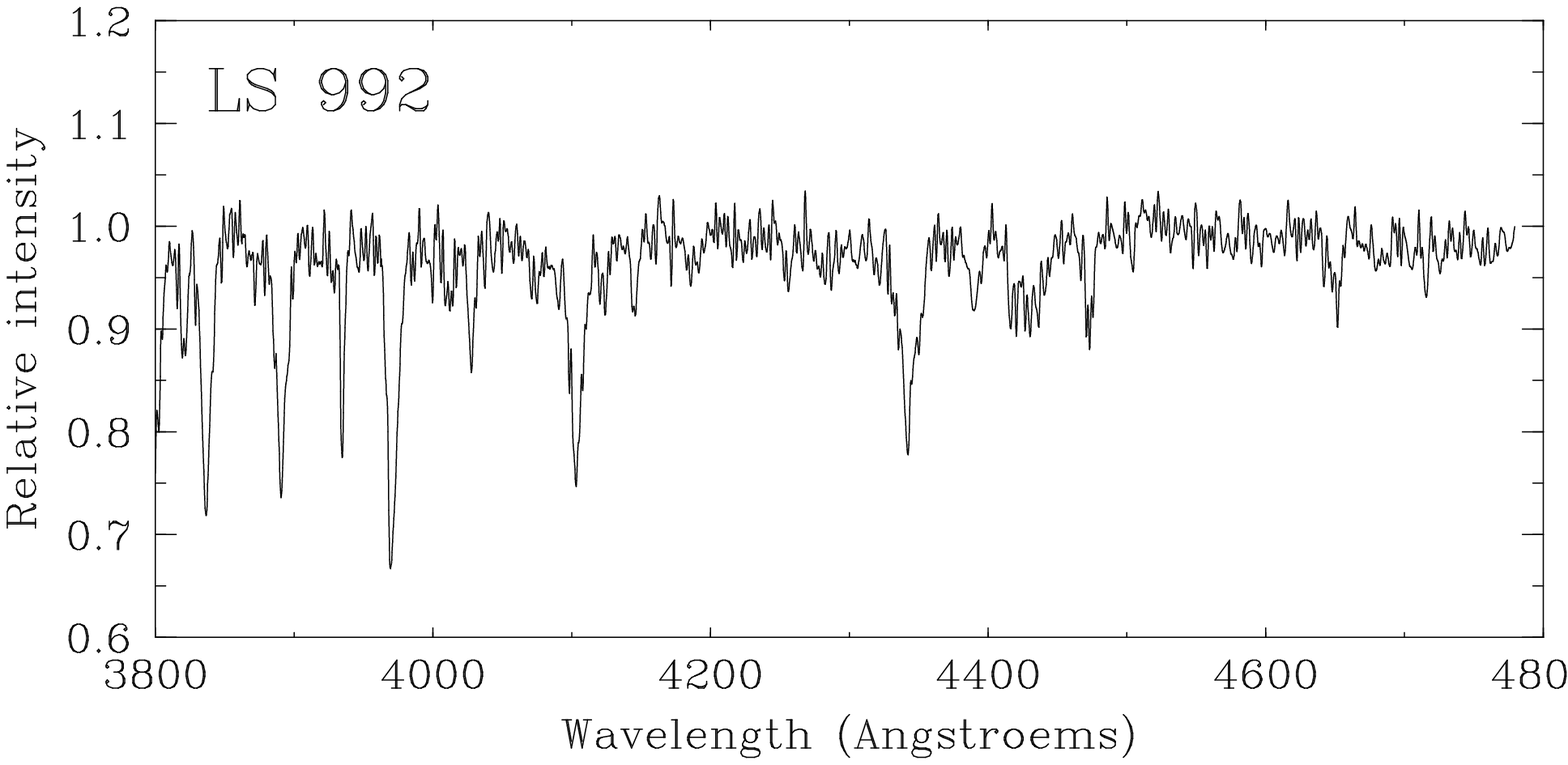,width=8.8cm,bbllx=1.5cm,bburx=23cm,bblly=2.5cm,bbury=13cm}
\caption[]{Rectified blue medium resolution spectrum of LS 992 obtained with
the ESO-MPI 2.2\,m + EFOSC2 on 1992 April 18. Exposure time was 15 min}
\label{spblueLS992}
\end{figure}

LS 992 is one of the X-ray brightest and hardest new Be/X-ray candidates in the ROSAT
all-sky survey.  Close inspection of B and I images (see Fig.~\ref{imageLS992}) failed to
reveal any alternative optical counterpart to the X-ray source. Optical spectroscopy of
object B, the second brightest in the error circle, suggests a late type star with
\Halpha \ and \Hbeta \ in absorption, leaving the B star as the most likely optical
counterpart of  RX J0812.4$-$3114.  Our blue medium resolution spectrum of LS 992 is
unfortunately somewhat too noisy to derive a very accurate spectral type.  However, the
\ion{Si}{III/IV} and \ion{He}{I} lines indicate a B0-1 spectral type and a
luminosity class V to III (see Fig.~\ref{spblueLS992}). The red spectra show the presence
of \Halpha \ emission with a central absorption core (see Fig. \ref{spredLS992}) similar
to that observed from LS 1698 (see below). The total \Halpha \ equivalent width does not
vary much over the one week interval between the two spectroscopic observations
(EW(\Halpha ) = $-$4.6 $\pm$ 0.1; $-$4.9 $\pm$ 0.1 on 1992 April 18 and 25 respectively).
However, the V/R ratio may have slightly evolved.

Optical photoelectric photometry by Reed (1990) gives V = 12.42; B-V = 0.41 and U-B  =
$-$0.69 in agreement with the B0-1 V-IIIe spectral type proposed. Assuming (B-V)$_{0}$ =
$-$0.28 yields E(B-V) = 0.69, consistent with that derived from the depth of the
$\lambda$ 4430 and $\lambda$ 6284 interstellar bands which imply E(B-V) = 0.7-1.0 and
E(B-V) = 0.8 $\pm$ 0.2 respectively.  

\begin{figure}
\psfig{figure=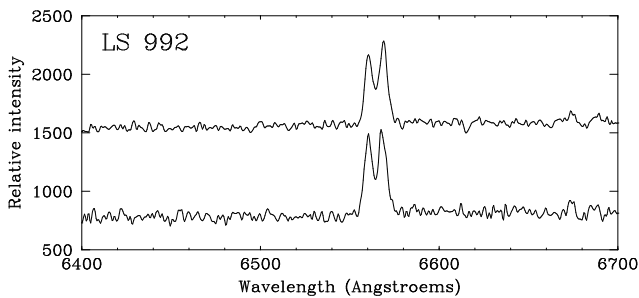,width=8.8cm,bbllx=1.5cm,bburx=23cm,bblly=2.5cm,bbury=13cm}
\caption[]{Red medium resolution spectrum of LS 992 showing the \Halpha \ emission.
The upper spectrum was exposed 20 min on 1992 April 18 and the lower one 5\,min on April 25. Both
spectra were obtained with the ESO-MPI 2.2\,m telescope}
\label{spredLS992}
\end{figure}

\begin{figure}
\psfig{figure=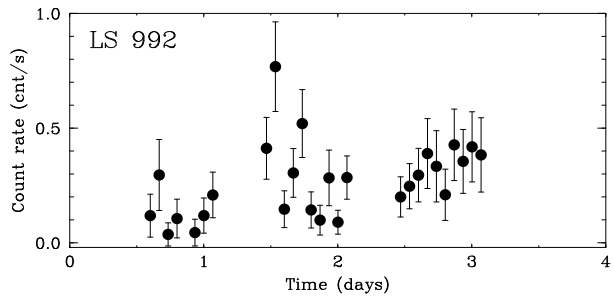,width=8.8cm,bbllx=1.5cm,bburx=23cm,bblly=2.5cm,bbury=13cm}
\caption[]{Survey light curve of RX J0812.4$-$3114, the source associated with LS 992.
Each data point corresponds to one satellite scan (10 to 32\,s every 96\,min)
integrated between 0.4 and 2.4\,keV. Time 0 is JD 2448192.5759}
\label{surveylcLS992}
\end{figure}

Although LS 992 is the second brightest X-ray binary candidate from the survey data,
the statistics are still not good enough to really constrain the shape of the X-ray
energy distribution. The spectrum is undoubtedly much harder (HR2 = +0.74$\pm$0.09)
than normal OB star emission.  Fitting simple power law models and fixing the column
density at its interstellar value (\nh \ = 5 10$^{21}$\,cm$^{-2}$) implies a photon
index in the range of +0.5 to $-$1.5 (95\% confidence level), well within the range
observed from accreting neutron stars in massive X-ray binaries (White et al. 1983).
The hard X-ray light curve plotted on Fig.~\ref{surveylcLS992} shows some evidence for
variability.
 
A long follow-up pointed PSPC observation on 1992 November 20 failed to recover the
source. A 3\,$\sigma$ upper limit of $\approx$  3 10$^{-3}$ \cnts\ can be set to the
count rate of RX J0812.4$-$3114, i.e. about 100 times fainter than during survey
observations.  Contemporaneous optical spectroscopy obtained at LNA, Brazil, on November
28 shows \Halpha \ emission with an EW = $-$4.6\,\AA , similar to that observed 7 months
before at La Silla.

No catalogued X-ray source is found at the position of LS 992. The large long term flux
variability, similar to that exhibited by many Be/X-ray transients and the
X-ray hardness strongly suggest that LS 992 is a new accreting Be/X-ray
system.

\subsubsection{LS 1698 (Group 1)}

\begin{figure}
\psfig{figure=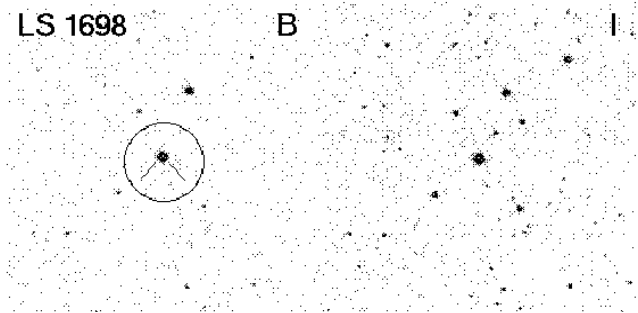,width=8.8cm,bbllx=3.5cm,bburx=18cm,bblly=10.5cm,bbury=18cm,clip=true}
\caption[]{B and I CCD images of the field of LS 1698 obtained on 1992 April 17 with EFOSC2 and
the ESO-MPI 2.2\,m telescope. Both images have 2\,min exposure time.
North is at top and East to the left. Each frame is $2\farcm83$$\times$$2\farcm83$ wide.
On the B image we plot the ROSAT pointing
90\% confidence error circle and show the position of LS 1698}
\label{imageLS1698}
\end{figure}

\begin{figure}
\psfig{figure=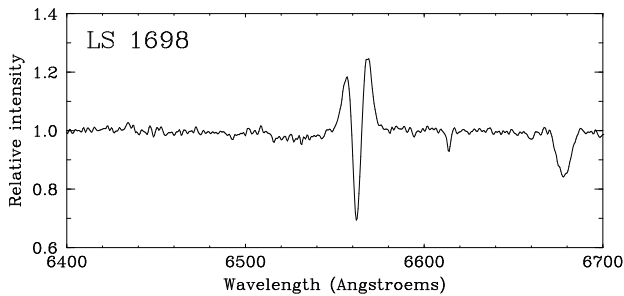,width=8.8cm,bbllx=1.5cm,bburx=23cm,bblly=2.5cm,bbury=13cm}
\caption[]{Rectified red medium resolution spectrum of LS 1698 showing the \Halpha \ emission.
This spectrum (15 min exposure time) is the sum of two spectra obtained on
1992 April 15 and 17 with the ESO-MPI 2.2\,m telescope}
\label{spredLS1698}
\end{figure}

\begin{figure}
\psfig{figure=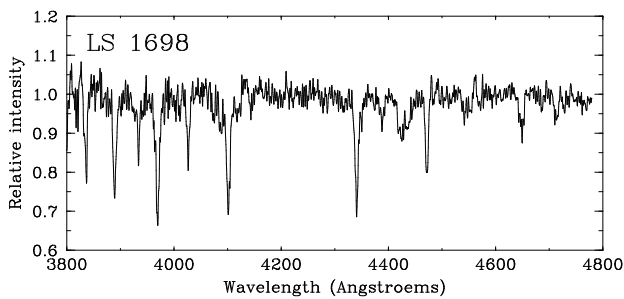,width=8.8cm,bbllx=1.5cm,bburx=23cm,bblly=2.5cm,bbury=13cm}
\caption[]{Rectified blue medium resolution spectrum of LS 1698 obtained with
the ESO-MPI 2.2\,m + EFOSC2 on 1992 April 17. Exposure time was 10\,min}
\label{spblueLS1698}
\end{figure}

Comparison of the I and B band CCD images failed to reveal any other bright or very red
star in the error circle which could be an alternative active corona identification (see
Fig.~\ref{imageLS1698}).  In the red, \Halpha \ is clearly seen in emission with a strong
absorption core (see Fig.~\ref{spredLS1698}). The same \Halpha \ profile was again
observed in February 1994. Our medium resolution blue spectrum shown in
Fig.~\ref{spblueLS1698} indicates a B0 V-IIIe type star.  The strength of the $\lambda$
4430 and $\lambda$ 6284 interstellar bands suggests a rather large interstellar
absorption E(B-V) = 0.75 $\pm$ 0.25. Using the GSC V magnitude and taking into account
the reddening estimated from interstellar bands lowers the distance of LS 1698 from
18\,kpc to $\approx$ 5\,kpc.  

The source was again detected during a dedicated pointed observation with a count rate
about 20 times lower. The survey energy distribution characterized by HR2 is hard and
HR1 is consistent with the large interstellar absorption. From these evidences, we
conclude that LS 1698 is most probably a new Be/X-ray system.

We note that LS 1698 is the likely optical counterpart of the hard X-ray transient
source 4U1036$-$56/3A 1036$-$565. Its position lies well inside the error box of the
Uhuru source and close to the Ariel V position and the ROSAT survey source is the only
one in the Uhuru and Ariel error boxes. The former identification of 3A 1036$-$565
with the bright Be star HD 91188 (Buckley et al. 1985) is thus probably not valid
since the star lies quite outside the error box and is furthermore not detected in the
ROSAT all-sky survey.  

The source was first observed by the Uhuru satellite (Forman et al. 1978) and OSO 7
(Markert et al. 1979).  Ariel V observed a flare in November 1974 (Warwick et al. 
1981). At maximum flare the flux was $\approx$ 2.4 10$^{-10}$\,\ergcm \ (2-10\,keV)
while the mean Uhuru flux was $\approx$ 1.0 10$^{-10}$\,\ergcm \ (2-10\,keV).  During
the survey, ROSAT detected a flux about 10 times less than during the years 1970-1976.

\subsubsection{LS 5039 (Group 1)}

\begin{figure}
\psfig{figure=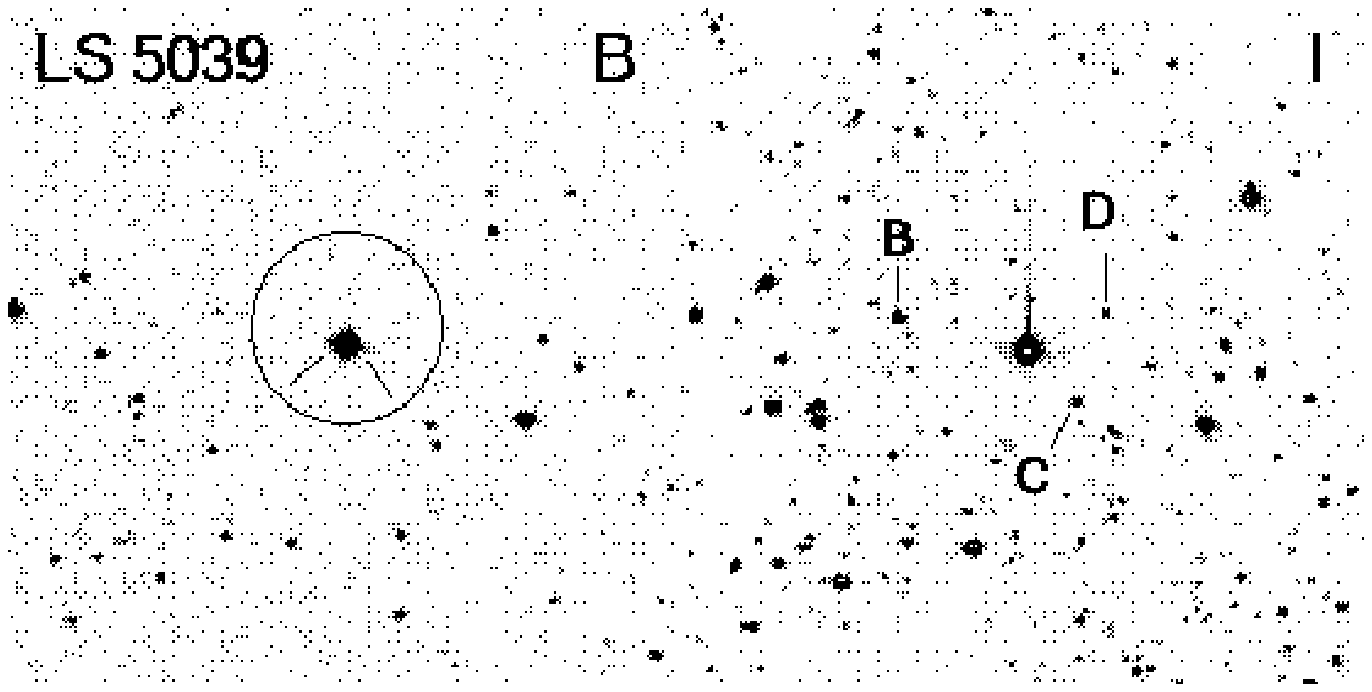,width=8.8cm,bbllx=3.5cm,bburx=18cm,bblly=10.5cm,bbury=18cm,clip=true}
\caption[]{B and I CCD images of the field of LS 5039 obtained on 1992 April 19 with EFOSC2 and
the ESO-MPI 2.2\,m telescope. Both images have 2\,min exposure time.
North is at top and East to the left.  Each frame is $2\farcm83$$\times$$2\farcm83$ wide. On the B image we plot t
he ROSAT survey
90\% confidence error circle and show the position of LS 5039. On the I band image we show the
position of the alternative optical candidates investigated spectroscopically and
apparently unrelated to the X-ray source}
\label{imageLS5039}
\end{figure}

From our B and I band images of the field of LS 5039 we could select 3 alternative
candidate stars designated B,C and D on Fig.~\ref{imageLS5039}. Low resolution
spectroscopic observations of these stars failed to reveal spectral signatures of any
known class of galactic or extragalactic X-ray sources.  All three candidates exhibit
\Halpha \ in absorption. Object B is an early M star.  However, it lacks the Balmer
emission which is nearly always found in X-ray active M stars (see e.g. Fleming et al.
1988; Motch et al. 1996b). Object C and D are too faint to be easily classified.
Therefore, LS 5039 constitutes the most likely optical counterpart of the X-ray source.

\begin{figure}
\psfig{figure=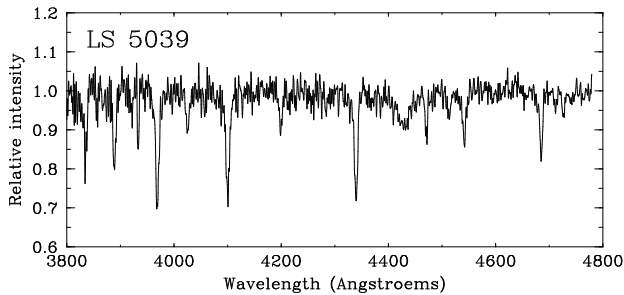,width=8.8cm,bbllx=1.5cm,bburx=23cm,bblly=2.5cm,bbury=13cm}
\caption[]{Rectified blue medium resolution spectrum of LS 5039 obtained with
the ESO-MPI 2.2\,m + EFOSC2 on 1992 April 19. Exposure time was 5\,min}
\label{spblueLS5039}
\end{figure}

The medium resolution blue spectrum of LS 5039 shown on Fig.~\ref{spblueLS5039} is
typical of an O7V star with well marked \ion{He}{II} \ $\lambda$ 4686 absorption and some
evidence for weak \ion{N}{III}\ $\lambda$ 4634-4642 emission, i.e.  an O7V((f))
classification. No \Halpha \ emission is present in our medium resolution red spectrum.

The optical photometry reported by Drilling (1975) gives V = 11.23, B-V  = 0.94, U-B =
$-$0.16.  A (B-V)$_{0}$ = $-$0.32 implies E(B-V) = 1.26, a value slightly larger than that
derived from the $\lambda$ 4430 and $\lambda$ 6284 interstellar band which both point at
E(B-V) = 0.8 $\pm$ 0.2.

During survey observations the X-ray source was one of the hardest of the OB/X-ray
correlation list. The source was detected again during a subsequent PSPC pointing.
The X-ray position derived from this observation is completely consistent with the
survey position. The count rate is close to that of the survey but owing to the
longer exposure time the slightly improved statistics allows to confirm the hardness
of the source, clearly compatible with an accreting neutron star or black hole seen
through the large intervening column density.
  
In spite of the slight increase of the bolometric correction with respect to the
automatic process which assumed a B0 III type, the updated \Lx/\Lbol \ ratio remains high
(10$^{-5}$), exceeding by an order of magnitude at least that expected from normal O
stars. We conclude that LS 5039 is a likely massive X-ray binary in which the compact
object may accrete from the O star high velocity wind.  

\subsubsection{HD 161103 (Group 3)} 

HD 161103 is a relatively bright and variable (V = 8.4-8.7) B2 V-IIIe star. 
Spectroscopic observations on 1994 February 13 show prominent Balmer emission with 
EW(\Halpha) = $-$32\,\AA . The GSC star nearest (29\arcsec ) to the B star (GSC
0683600952; V$\approx$ 12.4) but located slightly outside the 90\% confidence 
survey error circle was also observed at this occasion and failed to reveal
the signature characteristic of an active corona.  

The source was detected again during a short dedicated PSPC pointing with a count rate
about half of that recorded during the survey. The position resulting from the pointed
observation is now only 6\arcsec \ away from HD 161103 and the energy distribution is
clearly hard (HR2 = +0.47$\pm$0.21) although we still have too few photons to really
characterize the energy distribution. The \Lx/\Lbol \ ratio is close to the end of the
observed distribution for O stars (Sciortino et al.  1990).  However, this is still
$\approx$ 50 times more than for the B1-B2 type stars measured by Cassinelli et al.
(1994).  The corresponding un-absorbed 0.1-2.4\,keV X-ray luminosities are close to
10$^{32}$\,\ergs.

Our conclusion is that the evidences in favour of an accreting component are strong but
not firm enough in order to definitively claim that HD 161103 is a new Be/X-ray system.
However, this object is a good candidate and would certainly deserve follow-up X-ray
observations in a harder X-ray band.  

\subsubsection{SAO 49725 (Group 3)}

\begin{figure}
\psfig{figure=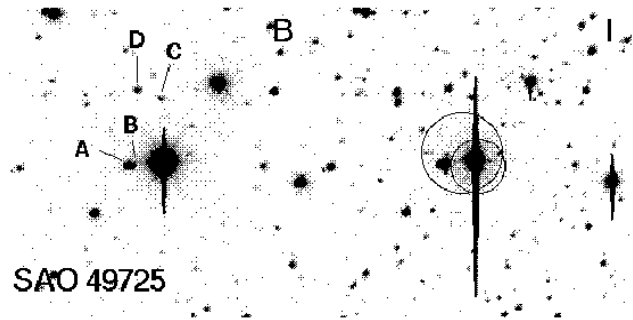,width=8.8cm,bbllx=3.5cm,bburx=18cm,bblly=10.5cm,bbury=18cm,clip=true}
\caption[]{B and I CCD images of the field of SAO 49725 obtained on 1992 May 27 with the CCD camera
at the OHP 1.2\,m telescope.
Both images have 10\,min exposure time.
North is at top and East to the left. Each frame is 3\farcm54$\times$3\farcm54 wide.
On the I image we plot the ROSAT survey and pointed
90\% confidence error circles. On the B band image we show the
position of the other possible optical candidates which were spectroscopically investigated}
\label{imageSAO49725}
\end{figure}

SAO 49725 is a relatively bright B0e star (V = 9.23) suffering a reddening E(B-V)
$\approx$ 0.67.  The survey source was recovered in a subsequent pointing with a count
rate of about twice of that given by the EXSAS analysis for the survey. Both survey
and pointed positions encompass SAO 49725 (see Fig.~\ref{imageSAO49725}).  

\begin{figure}
\psfig{figure=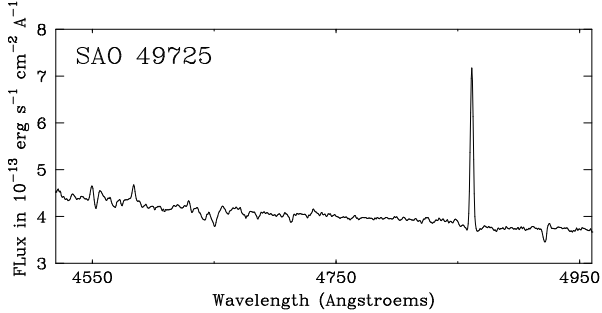,width=8.8cm,bbllx=1.5cm,bburx=23cm,bblly=2.5cm,bbury=13cm}
\caption[]{Blue medium resolution spectrum of SAO 49725 obtained on 1992 October 22 at OHP
with the 1.93\,m telescope and the CARELEC spectrograph. Exposure time is 20\,min}
\label{SAO49725spblue}
\end{figure}

None of the red excess candidates located in or close to the X-ray error circles and
marked on Fig.~\ref{imageSAO49725} display any signature of chromospheric activity. 
Objects A, D and C are late G-F stars. Object B is a M star without any detectable
Balmer or \ion{Ca}{II}\  emission and is therefore an unlikely optical counterpart.  Our
blue medium resolution spectrum (see Fig.~\ref{SAO49725spblue}) displays, in addition to
\Hbeta , \ion{O}{II}, \ion{C}{III}\ and weak \ion{He}{II}\  lines which indicate a B0.5
type in agreement with former determinations. The relative strength of the \ion{He}{II}\
$\lambda$ 4686 and \ion{He}{I}\ $\lambda$ 4713 and that of the \ion{Si}{II}\ lines
suggests a dwarf or giant luminosity class.  

\begin{figure}
\psfig{figure=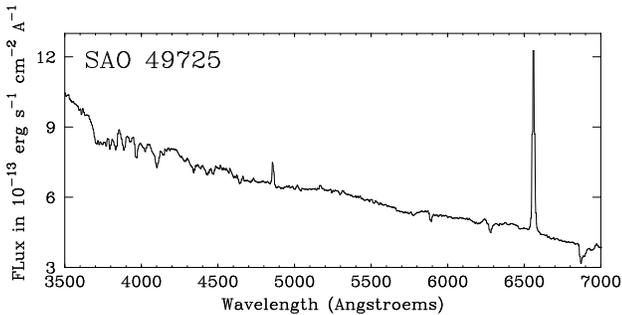,width=8.8cm,bbllx=1.5cm,bburx=23cm,bblly=2.5cm,bbury=13cm}
\caption[]{Low resolution spectrum of SAO 49725 obtained on 1991 June 8 at OHP
with the 1.93\,m telescope and the CARELEC spectrograph. The 60\% flux difference
between this spectrum and that shown in Fig.~\ref{SAO49725spblue} is due
to inaccurate flux calibration. Exposure time is 2\,min}
\label{SAO49725spred}
\end{figure}

Fig.~\ref{SAO49725spred} shows the prominent \Halpha \ (EW = $-$30 $\pm$ 1\,\AA ) and
\Hbeta \ (EW = $-$2.0 $\pm$ 0.1\,\AA ) emission lines. The Balmer continuum is also
seen in emission revealing the presence of a large circumstellar envelope.  

The pointing data show that the X-ray source is relatively hard.  With the updated
optical and X-ray data, the \Lx/\Lbol \ ratio remains close to 3 $10^{-6}$. This is
rather strong evidence, although not fully compelling, for an accreting compact object
around SAO 49725.

\subsection{Uncertain cases}

\hskip 0.5cm 
HD 38087 (Group 1) : HD 38087 is a B5V star (Schild \& Chaffee 1971) located in the
Orion OB1 association.  The distance resulting from the photometry is 470\,pc in
agreement with that derived by Warren \& Hesser (1978) for the belt subregion ($\sim$
440\,pc). However, the star is associated with a reflection nebulosity and may have an
anomalous reddening law (R$_{\rm V}$\ = \ A$_{\rm V}$/E(B-V) = 5.3-5.7; Whittet \& van
Breda 1980, Cardelli et al.  1989) possibly caused by unusually large grains (Snow \&
Witt 1989). These authors derive a total \nh = 4.0 $\pm$ 0.5 $10^{21}$\,cm$^{-2}$
whereas with an observed E(B-V) = 0.29 the expected \nh \ should be $\approx$ 2.3
$10^{21}$\,cm$^{-2}$.  

The position of the source was covered by two ROSAT pointings and it was detected on
each occasion with a mean count rate similar to that observed during survey
observations.  Survey and pointing spectra have rather soft HR2s usually not observed
in OB/X-ray systems.  The spectrum accumulated from WG900386 can be represented by a
two temperature thin thermal Raymond Smith spectrum (kT1 = 0.28\,keV; kT2 = 3.1\,keV;
\nh = 3 10$^{21}$\,cm$^{-2}$) typical for active coronae.  Assuming a distance of
470\,pc the un-absorbed 0.1-2.4\,keV X-ray luminosity is $\approx$ 7 10$^{31}$\,\ergs
(two temperature plasma), close to that computed by the automatic process. Such a high
X-ray luminosity is sometimes observed during flaring states in the most active T
Tauri stars (see e.g. Montmerle et al. 1983).  With an estimated cluster age of 5
10$^{6}$ yr for the OB1 b1 group (Warren \& Hesser 1978) and considering the softness
of the source, the X-ray emission could still originate from a very active late type
star physically linked to HD 38037. The emission line star LkHA 291 is apparently
located well outside the ROSAT error circle as seen from the finding chart in
Herbig and Kuhi (1963).

HD 53339 (Group 1) : This B3V star belongs to the Canis Major OB1 association located
at a distance of $\approx$ 1150\,pc (Claria 1974). It consists of a close pair
separated by 1\farcs1 and with a magnitude difference of 0.9$^{\rm m}$. If the pair is
physically linked, the faint component is a late B star with little intrinsic X-ray
emission. The Einstein IPC did not detect X-ray emission from the region with an upper
limit of 5 10$^{-3}$ \cnts \ probably consistent with the PSPC count rate of 2.3
10$^{-2}$ \cnts . Hardness ratios are marred by large errors and compatible with both
an active corona or a hard accreting source. At the time of writing this paper the AO-3
pointing data covering HD 53339 were not available from the archive.

\begin{figure}
\psfig{figure=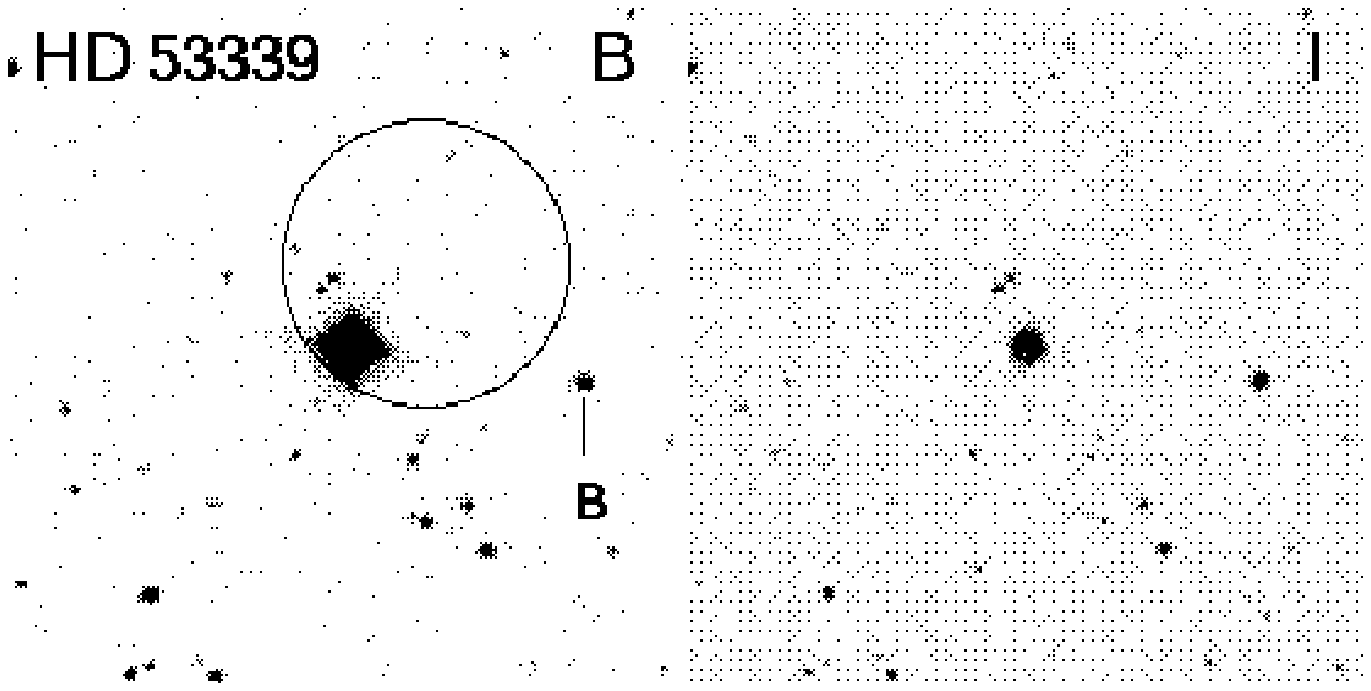,width=8.8cm,bbllx=3.5cm,bburx=18cm,bblly=10.5cm,bbury=18cm,clip=true}
\caption[]{B and I CCD images of the field of HD 53339 obtained on 1992 April 22 with EFOSC2 and
the ESO-MPI 2.2\,m telescope. Both images have 5\,min exposure time.
North is at top and East to the left. Each frame is $2\farcm83$$\times$$2\farcm83$ wide. On the B image we plot th
e ROSAT survey
90\% confidence error circle and show the position of a late K-M star which we
have investigated spectroscopically and which does not exhibit \Halpha \ emission}
\label{imageHD53339}
\end{figure}

Our optical observations failed to reveal any alternative bright optical counterpart. 
However, several faint stars in the error circle (see Fig.~\ref{imageHD53339}) would
deserve further investigation before settling the case. The reddish object B is a late
K-M star without \Halpha \  emission. With an estimated un-absorbed 0.1-2.4\,keV X-ray
luminosity of $\approx$ 6.5 10$^{31}$\,\ergs \ (computed using the results of the EXSAS
analysis of the survey), HD 53339 clearly deserves further optical and X-ray monitoring.

HR 2875 (Group 1) : HR 2875 is a bright (V = 5.41) B5Vp star also detected by the ROSAT
Wide Field Camera (Pounds et al. 1993). The PSPC energy distribution is extremely soft
and therefore the standard process computing the \Lx/\Lbol \ ratio, which assumed a
10$^{7}$\,K thin thermal emission, largely overestimated the X-ray flux in that
particular case. No photon is detected above 350\,eV in the ROSAT survey data and the
overall PSPC spectrum looks like that of an isolated hot white dwarf. Considering the
very low probability of a positional coincidence with such a bright star, the
hypothetical white dwarf should be physically linked to the B5 star and at a distance
of 190\,pc the observed count rate is compatible with this hypothesis (see e.g. Barstow
et al. 1994).  The expected `normal' harder X-ray emission from the early type star
should give a count rate $\approx$ 200 times smaller than observed and virtually be
undetectable in our data.

LS IV -12 70 (Group 1) : A source located at a position consistent with that found in
the survey was re-detected during the PSPC pointing WG 500040. However, the pointed
position is incompatible with the B star and a red excess object now lies in the
middle of the revised error circle.  

HD 38023 (Group 2) : This position was not optically investigated.  The B4V star HD
38023 is located in Orion inside a reflection nebula. The X-ray source was detected
again during a long PSPC pointing with a count rate similar to that recorded during the
survey. HD 38023 now falls within the 90\% confidence radius of the pointed position. 
It appears as the brightest spot of a patch of several nearby sources. RX
J0542.3$-$0807 is variable between 0.01 and 0.04 \cnts\ on a time scale of two days and
displays a soft energy distribution.  Fixing the interstellar absorption to the value
derived from the optical colours (\nh = 3.7 10$^{21}$\,cm$^{-2}$) a thin thermal
Raymond-Smith spectrum fit yields kT = 0.85 $\pm$ 0.12\,keV indicating that the PSPC
energy distribution is compatible with that expected from an active corona.  The
corresponding mean un-absorbed 0.1-2.4\,keV X-ray luminosity of $\approx$ 1.7
10$^{31}$\,\ergs \ is consistent with the luminosities derived by running the automatic
process on survey data. Considering the X-ray luminosity and softness of the source the
odds are more in favour of a young late type star physically related to HD 38023 rather
than for a genuine accreting binary.

LS V +29 14 (Group 2) : This position was not optically investigated. Comparison of POSS
O and E plates reveals the presence close to the survey X-ray position of at least two
faint red candidates which could be Me stars. 

LS 5038 (Group 2) : The field of LS 5038 was covered by a long PSPC pointed
observation. A weak source (1.8 10$^{-3}$ \cnts , i.e. about 10 times fainter than the
survey count rate) is detected at a position probably compatible with the survey
position. The source detected during the pointing is now farther away (93\arcsec) from
LS 5038 than was the survey position ruling out an association with the early type
star.  The pointed position is closer (69\arcsec) to the bright (V = 8.0) K0IV star HD
169651, however, the late type star is still outside the 90\% confidence radius.  Our
optical spectroscopic observations failed to identify any active corona in the close
neighbourhood of the early type star and revealed that LS 5038 is a Be star (EW(\Halpha
) = $-$4.4 $\pm$ 0.3\,\AA ) exhibiting a central \Halpha \ absorption core similar to
that seen in LS 1698 and LS 992. The $\lambda$ 6284 interstellar band indicates E(B-V)
$\approx$ 0.85. Our low resolution spectroscopic data do not allow accurate spectral
type determination. We cannot completely rule out the unlikely possibility that we have
witnessed a hard X-ray flare from an accreting object around LS 5038 during the survey
observation.

HD 36262 (Group 3) : HD 36262 is another early type star in Orion displaying an X-ray
excess. With a B3V spectral type and an un-absorbed X-ray luminosity of 3
10$^{31}$\,\ergs (using EXSAS results), the most likely explanation of the X-ray excess
is the presence of an extremely active companion star. The hardness ratios are
compatible with a rather soft stellar source.  

\subsection{Likely non-accreting sources}

\subsubsection{LS III +46 11 (Group 1)}

\begin{figure}
\psfig{figure=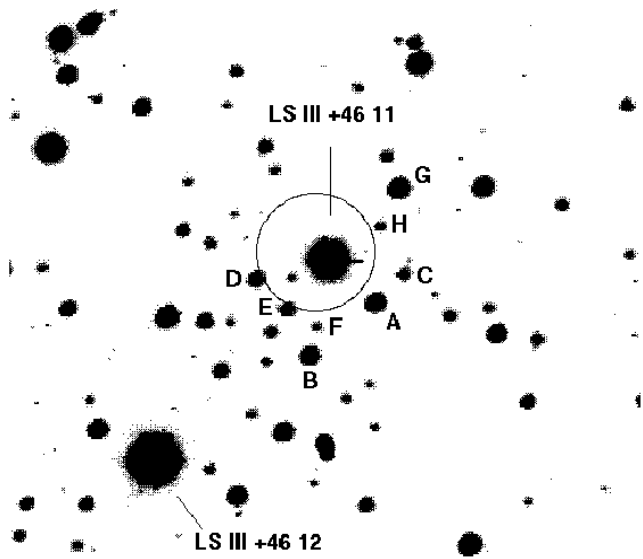,width=8.8cm,bbllx=3.5cm,bburx=18cm,bblly=7cm,bbury=20cm,clip=true}
\caption[]{B CCD image of the field of LS III +46 11 obtained on 1992 May 27 with the RCA3 CCD
camera and the 1.2\,m telescope at OHP. We show the 90\% confidence ROSAT survey error
circle and the position of the alternative candidate stars for which we obtained
optical spectroscopy. North is at top and East to the left.
The field is 3\farcm54$\times$3\farcm54 wide}
\label{imageLSIII+4611}
\end{figure}

This OB star is a probable member of the Berkeley 90 cluster (Sanduleak 1974) and
located within the HII region Sharpless 115 (Harten \& Felli 1980). LS III +46 11 is
well placed within the ROSAT survey error circle. Spectroscopic observations of all
nearby objects marked on Fig.~\ref{imageLSIII+4611} failed to reveal any convincing
alternative optical counterpart leaving the OB star as the most likely identification
of RX J2035.2+4651.  

\begin{figure}
\psfig{figure=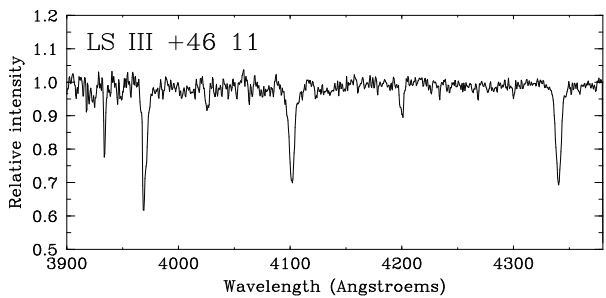,width=8.8cm,bbllx=1.5cm,bburx=23cm,bblly=2.5cm,bbury=13cm}
\caption[]{Rectified blue medium resolution spectrum of LS III +46 11 obtained with
the OHP 1.93\,m telescope and the CARELEC spectrograph. The spectrum shown here is the mean of
4 spectra collected in the time interval 1991 November 20-24. Total exposure time is 100\,min}
\label{spblueLSIII+4611}
\end{figure}

\begin{figure}
\psfig{figure=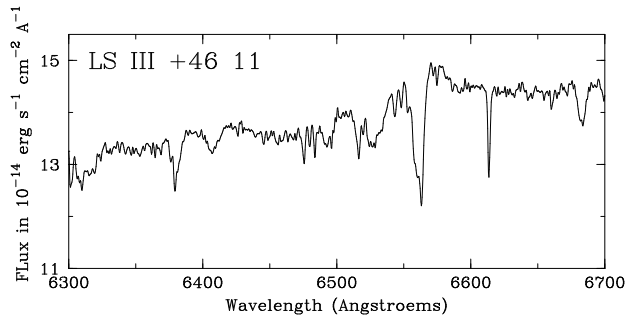,width=8.8cm,bbllx=1.5cm,bburx=23cm,bblly=2.5cm,bbury=13cm}
\caption[]{Red medium resolution spectrum of LS III +46 11 obtained with
the OHP 1.93\,m telescope and the CARELEC spectrograph on 1992 December 15.
A broad \Halpha \ component (FWHM $\approx$ 60\,\AA) is visible and indicative of the
high stellar luminosity}
\label{spredLSIII+4611}
\end{figure}

Medium resolution (\lam \lam 3900-4400\,\AA ; Fig. ~\ref{spblueLSIII+4611} and \lam \lam
6300-6700\,\AA ; Fig. ~\ref{spredLSIII+4611}) and low resolution (\lam \lam
3800-7100\,\AA ; not shown) optical spectroscopy indicate a very hot giant star. The
\ion{He}{II} $\lambda$ 4541 / \ion{He}{I} $\lambda$ 4471 and the \ion{He}{II}  $\lambda$
4025 / \ion{He}{II}  $\lambda$ 4200 line ratios both indicate an O3-O5 spectral type
(see Fig.~\ref{spblueLSIII+4611}).  The probable presence of \ion{N}{IV}\ $\lambda$ 4058
emission also visible in our low resolution spectrum and possible \ion{N}{V} $\lambda$
4606-4620 absorption suggest a class III luminosity.  Broad \Halpha \ emission is
conspicuous and indicative of the high luminosity of the star (see
Fig.~\ref{spredLSIII+4611}).  We thus propose an O3-O5III(f)e spectral type for LS III
+46 11.

In a subsequent dedicated pointed observation LS III +46 11 was again detected with a
count rate consistent with that recorded during the survey (0.046 \cnts ). The
nearby O6 star LS III +46 12 was also detected with a count rate of 0.016 $\pm$ 0.004
\cnts .

In this particular case, the assumption of a default B0 III spectral type by the
automatic SIMBAD/ROSAT correlation process led to a clear overestimation of the
\Lx/\Lbol \ ratio.  Using an O3 III spectral type now gives  \Lx /\Lbol $\approx$ 2.7
10$^{-6}$. With this spectral type the star is at 1.9\,kpc, in agreement with the
distance of the Be 90 cluster and the 0.1-2.4\,keV un-absorbed luminosity is of the
order of 5 10$^{33}$\,\ergs . We also note that the ratio of the PSPC count rate from
LS III +46 11 over that of  LS III +46 12 is similar to the ratio of their bolometric
luminosities.  This implies that most of the ionizing power in S 115 is actually
provided by  LS III +46 11 rather than by LS III +46 12.  Although extremely rare,
such high X-ray to bolometric ratios and soft X-ray luminosities have been reported
for a few O type stars (Sciortino et al. 1990).  We therefore conclude that there is
probably no need for an accreting component in order to explain the X-ray emission of
LS + 46 11.  However, considering the extreme value of the X-ray luminosity this star
would certainly deserve further detailed investigation.

\subsubsection{Other sources}

\begin{figure}
\psfig{figure=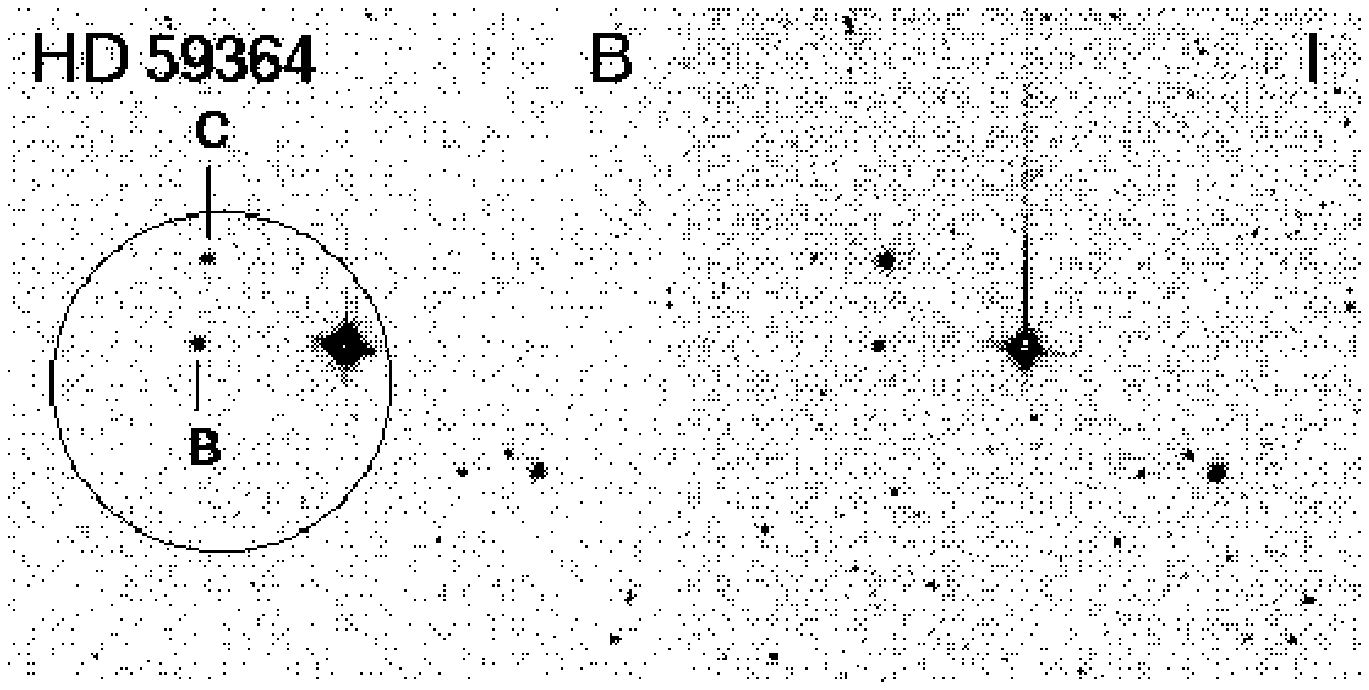,width=8.8cm,bbllx=3.5cm,bburx=18cm,bblly=10.5cm,bbury=18cm,clip=true}
\caption[]{B and I CCD images of the field of HD 59364 obtained on 1992 April 18 with EFOSC2 and
the ESO-MPI 2.2\,m telescope. Both images have 2\,min exposure time.
North is at top and East to the left.  Each frame is $2\farcm83$$\times$$2\farcm83$ wide. On the B image we plot t
he ROSAT survey
90\% confidence error circle and show the position of the objects investigated
spectroscopically. We identify the X-ray source with the red Me star C}
\label{imageHD59364}
\end{figure}

\begin{figure}
\psfig{figure=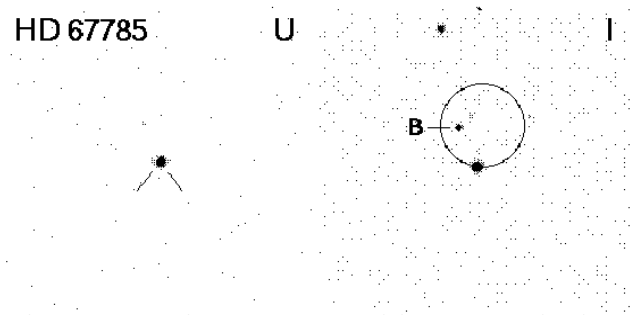,width=8.8cm,bbllx=3.5cm,bburx=18cm,bblly=10.5cm,bbury=18cm,clip=true}
\caption[]{U and I CCD images of the field of HD 67785 obtained on 1992 April 17 with EFOSC2 and
the ESO-MPI 2.2\,m telescope. Both images have 30\,s exposure time.
North is at top and East to the left. Each frame is $2\farcm83$$\times$$2\farcm83$ wide. On the U image we show th
e position of HD 67785.
On the I image we plot the ROSAT survey
90\% confidence error circle and show the position of object B which is the
actual optical counterpart of RX J0807.2$-$5053}
\label{imageHD67785}
\end{figure}

\begin{figure}
\psfig{figure=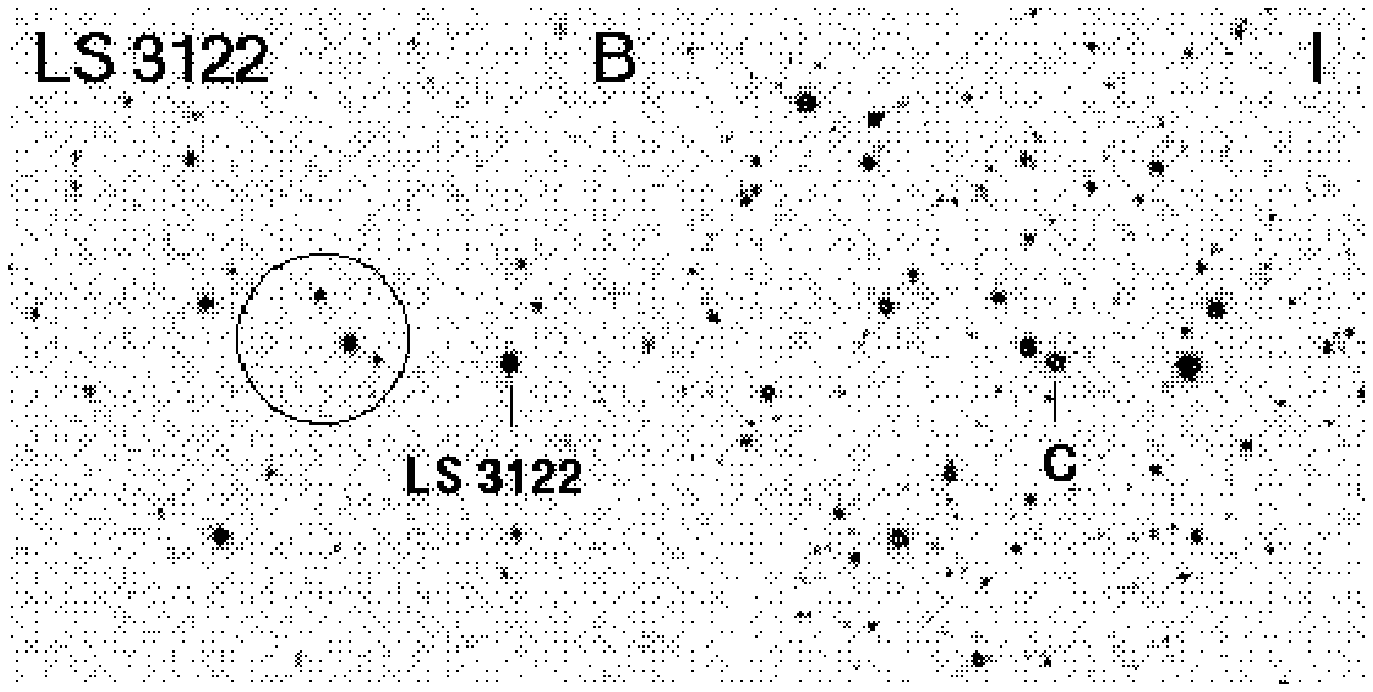,width=8.8cm,bbllx=3.5cm,bburx=18cm,bblly=10.5cm,bbury=18cm,clip=true}
\caption[]{B and I CCD images of the field of LS 3122 obtained on 1992 April 17 with EFOSC2 and
the ESO-MPI 2.2\,m telescope. Both images have 2\,min exposure time.
North is at  top and East to the left. Each frame is $2\farcm83$$\times$$2\farcm83$ wide. On the B image we plot t
he ROSAT survey
90\% confidence error circle and show the position of LS 3122. On the I band image we show the
position of object C, the Me star optical counterpart of RX J1335.5$-$6211}
\label{imageLS3122}
\end{figure}

\hskip 0.5cm 
HD 59364 (Group 1): RX J0728.6$-$2629 is probably identified with a red Me star
(object C in Fig. \ref{imageHD59364}) which exhibits strong \ion{Ca}{II}\ H\&K emission
and Balmer emission.

HD 67785 (Group 1): RX J0807.2$-$5053 is most likely identified with object "B" (see
Fig.~\ref{imageHD67785}). This late K-M star displays \ion{Ca}{II}\ H\&K emission
consistent with the X-ray flux.
   
HD 165424 (Group 1): Follow-up pointed PSPC observations give an improved position only
3\arcsec \ away from the V = 9.0 K0 star HD 315203 and the former proposed B4II
identification HD 165424 is now outside the ROSAT error circle. Systematic attitude
errors are unlikely to be large for this pointing since we detect the X-ray emission of
the symbiotic star He 3 -1591 only 10\arcsec \ away from the optical position. We
conclude that RX J1806.8$-$2606 should rather be identified with the late type star HD
315203.

BD + 60 282 (Group 2) : RX J0136.7+6125 is probably identified with the V = 10.7 star
GSC 0403100953, 10\arcsec \ away from the ROSAT survey position. Optical spectroscopy
shows that GSC 0403100953 is a late type star displaying strong \ion{Ca}{II} \ H\&K
emission consistent with the measured X-ray flux.

LS I +61 298 (Group 2) : RX J0234.4+6147 is probably identified with the V = 14.1 red
excess object (GSC 0404700465) located 10\arcsec \ away from the survey position.  Red
medium resolution spectroscopy reveals the presence of narrow \Halpha \ emission
superposed on a deep absorption profile indicating that GSC 0404700465 is a late F or
G type active star.  The source was detected again during an AO3 pointing with a count
rate similar to that of the survey observation. The pointed position is closer to the
B star (19\arcsec) but still compatible with the emission line candidate star
(34\arcsec).

LS 3122 (Group 2) : The Me star (object C in Fig.~\ref{imageLS3122}) is the likely
counterpart of RX J1335.5$-$6211. The star is located well inside the 90\% confidence
error circle and medium resolution red spectroscopy revealed strong \Halpha \ emission
(EW $\approx$ 6.0\,\AA ).

\begin{figure}
\psfig{figure=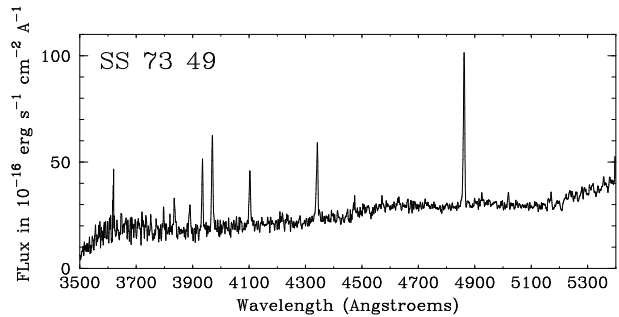,width=8.8cm,bbllx=1.5cm,bburx=23cm,bblly=2.5cm,bbury=13cm}
\caption[]{Flux calibrated blue medium resolution spectrum of SS 73 49,
the proposed optical counterpart of RX J1559.2$-$4157. The spectrum was obtained
on 1992 April 20 with EFOSC2 and the ESO-MPI 2.2\,m telescope. Exposure time is 10\,min}
\label{spblueSS7349}
\end{figure}

\begin{figure}
\psfig{figure=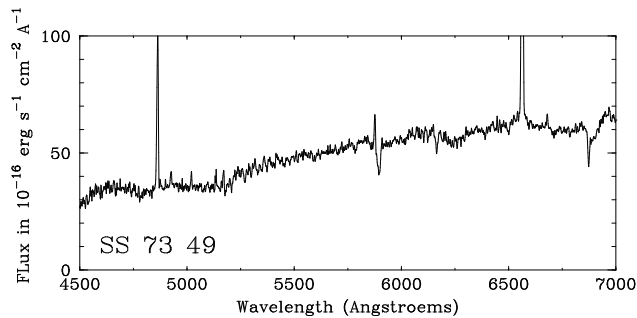,width=8.8cm,bbllx=1.5cm,bburx=23cm,bblly=2.5cm,bbury=13cm}
\caption[]{Flux calibrated low resolution spectrum of SS 73 49, the proposed optical
counterpart of RX J1559.2$-$4157. The spectrum was obtained
on 1992 April 20 with EFOSC2 and the ESO-MPI 2.2\,m telescope. Exposure time is 5\,min}
\label{spredSS7349}
\end{figure}

SS 73 49 (Group 2) : The emission line star Wray 15-1400 = SS 73 49 was discovered by
Wray (1966) and is identified with GSC 0784600316. Using objective prism low
resolution spectra Sanduleak \& Stephenson (1973) classify the object as a Be type
star.  Our medium (Fig.~\ref{spblueSS7349}) and low resolution
(Fig.~\ref{spredSS7349}) spectra show strong Balmer and \ion{Ca}{II}\  H\&K emission.
We classify SS 73 49 as a strong line T Tauri star rather than Be. \ion{Ca}{II}\ H\&K
emission is unusual in Be stars and there are evidences for Mg\,b and TiO $\lambda
\lambda$ 6187-6215 bands in absorption suggesting a late K star underlying continuum.
The \ion{Ca}{II}\ H\&K flux is consistent with that expected from the PSPC count rate
for active coronae.  The survey hardness ratios are also compatible with those
normally observed from young active stars.  Interactive analysis of the survey photons
yields a position slightly closer to the active star at the edge of the 95\%
confidence error circle. We conclude that RX J1559.2$-$4157 is optically identified
with the T Tauri star SS 73 49.

LS IV -05 35 (Group 2) : RX J1900.7$-$0503 is most probably identified with the V =
12.2 star GSC 0513601711 located within the 90\% confidence error circle. Optical
spectroscopy reveals noticeable \ion{Ca}{II} \ H\&K emission from GSC 0513601711.

\section{Discussion}

We summarize in Table \ref{summary1} the final list of identifications for all 24 sources
resulting from our follow-up optical/X-ray studies. Spectral types were updated using our
own determinations. In total we have discovered 5 very likely new massive X-ray binaries
with an additional 2 good candidates needing further confirmation.  Table \ref{summary2}
lists the main optical and X-ray characteristics of the 7 new massive X-ray binary
candidates. Distances, X-ray luminosity and luminosity ratios are subject to various
sources of error such as count rate statistics and unknown X-ray energy distribution,
inaccuracies in the interstellar absorption and absolute photometric calibrations.  A
typical error of 25\% to 50\% on all three quantities is probably realistic. The same
errors apply to data listed in Table \ref{knownxrbs}.
 
\subsection{Interlopers}

The cross-correlation in position of large catalogues with the unprecedented number of
sources present in the ROSAT all-sky survey unavoidably leads to numerous spurious
matches. Further selection of associations with extreme X-ray versus optical
characteristics such as the one performed here enhances the fraction of wrong
identifications and our present study clearly illustrates the need to ensure
identification using X-ray and especially optical follow-up observations.  

Several points of interest concerning the statistics of positional coincidence may be
noted.  First, all but 4 sources have a confirmed or a likely alternative optical
counterpart within the 90\% confidence radius. Among the 8 OB/X-ray associations
outside the SASS 95\% confidence radius (group 2) only one, SS 73 49 is confirmed. 
These evidences demonstrate the reliability of the survey ROSAT XRT positions and show
that the centering statistics is well enough understood at least phenomenologically to
be used as a constraining tool for systematic optical identification.  Second, 3
sources listed in group 1 have a confirmed alternative optical counterpart physically
unrelated to the OB star. A fourth case may be that of LS IV -12 70. This rate would
be close to the number of spurious matches within the 95\% confidence error radius
that was estimated in Section 2.5 on the basis of the size of the cross-correlated
samples. These three to four cases do appear in the first group consistent with the
expectation in Section 2.5 that spurious associations will be preferentially found
among the high \Lx/\Lbol \ objects. However, since similar comprehensive optical
follow-up has not been carried out for the remainder of the 108 OB/X-ray associations
no definite conclusions about the validity of the estimates for spurious matches can
be drawn.

\begin{table*}
  \caption{Summary of optical identifications. A.C. stands for active coronae.
Spectral types are from our own determination or from the literature.
Horizontal lines divide the three groups of candidates 
defined in Table 4}
    \label{summary1}
\begin{tabular}{lllll}
\hline
OB target    & ROSAT Source     & Optical         & Nature  & Spectral Type \\
Candidate    & Name             & Counterpart     &         &                \\
\hline 
LS I +61 235 & RX J0146.9+6121  & LS I +61 235    & Be/X    &  B1IIIe    \\           
BSD 24- 491  & RX J0440.9+4431  & BSD 24- 491     & Be/X    &  B0V-IIIe   \\ 
HD 38087     & RX J0542.9$-$0218  & HD 38087 ??     & A.C. ?? &            \\
HD 53339     & RX J0704.2$-$1123  & HD 53339 ??     & A.C. ?? &             \\
HD 59364     & RX J0728.6$-$2629  & Object C        & A.C.    &  Me        \\
HR 2875      & RX J0729.0$-$3848  & HR 2875 + WD ?? & WD ??   &             \\
HD 67785     & RX J0807.2$-$5053  & Object B        & A.C.    &  K-M          \\
LS 992       & RX J0812.4$-$3114  & LS 992          & Be/X    &  B0.5V-IIIe  \\
LS 1698      & RX J1037.5$-$5647  & LS 1698         & Be/X    &  B0V-IIIe \\
HD 165424    & RX J1806.8$-$2606  & HD 315203       & A.C.    &  K0V       \\
LS 5039      & RX J1826.2$-$1450  & LS 5039         & OB/X    &  O7V((f))    \\ 
LS IV -12 70 & RX J1830.7$-$1232  & ??              & A.C. ?? &            \\
LS III +46 11& RX J2035.2+4651    & LS III +46 11   & `Normal' O&  O3-O5 III(f)e \\
\hline
BD +60 282   & RX J0136.7+6125  & GSC 0403100953  & A.C.    &  G-K          \\
LS I +61 298 & RX J0234.4+6147  & GSC 0404700456  & A.C.    &  F-G ?        \\
HD 38023     & RX J0542.3$-$0807  & HD 38023 ??     & A.C. ?? &               \\
LS V +29 14  & RX J0553.0+2939  & ??              & ??      &               \\
LS 3122      & RX J1335.5$-$6211  & Object C        & A.C.    &  Me           \\
SS 73 49     & RX J1559.2$-$4157  & SS 73 49        & A.C.    &  T Tau         \\
LS 5038      & RX J1826.1$-$1321  & ??              & ??      &               \\
LS IV -05 35 & RX J1900.7$-$0503  & GSC 0513601711  & A.C.    & K-M           \\
\hline
HD 36262     & RX J0531.0+1206  & HD 36262  ??    & A.C. ?? &               \\  
HD 161103    & RX J1744.7$-$2713  & HD 161103       & Be/X ?? & B2 V-IIIe      \\
SAO 49725    & RX J2030.5+4751  & SAO 49725       & Be/X ?? & B0.5V-IIIe     \\
\end{tabular}
\end{table*}

\begin{table*} 
\caption{Optical and X-ray characteristics of the new OB/X-ray candidate systems. In this table we use the
revised spectral types derived from our optical observations and the maximum X-ray count rates measured from survey
or pointed PSPC observations and extracted using EXSAS. The last column lists the ratio of the maximum to minimum
count rate observed between survey and pointed observations when significant variability is detected. 
The first five objects can be considered as very good
OB/X-ray candidates whereas the last two require final confirmation of their X-ray excess} 

\label{summary2} % Herewe take maximum Lx from survey or pointing 
\begin{tabular}{llrllrllllr} 
\hline Optical      & Spectral &  V    & 
B-V  &  U-B  &  E(B-V)     &EW\Halpha & d    & \Lx/\Lbol & \Lx       & X-ray \\ 
Name        & Type     &       &    
&       &             & (\AA )   & (kpc) &           &(\ergs )  &Variability\\ 
\hline 
LS I +61 235 & B1IIIe   &11.33 &  0.82 & $-$0.39 &1.09         & $-$7 $-$10   & 2.9 & 2.7 10$^{-4}$ & 2.7 10$^{34}$ &1.8\\ 
BSD 24- 491  & B0V-IIIe &10.78 &  0.61 & $-$0.36 &0.90         & $-$2       & 3.2 & 2.2 10$^{-5}$ & 6.0 10$^{33}$ &2.5\\ 
LS 992       &B0.5V-IIIe&12.42 &  0.41 & $-$0.69 & 0.70        & $-$4.8     & 9.2 & 7.0 10$^{-4}$ & 1.3 10$^{35}$ &$\geq$95\\ 
LS 1698      & B0V-IIIe &11.3  &       &       &$\approx$0.75  & $-$1.5     & 5.0 & 3.9 10$^{-5}$ & 1.1 10$^{34}$ &20\\ 
LS 5039      & O7V((f)) &11.23 &  0.94 & $-$0.16 & 1.26        & +2.7     & 3.1 & 1.0 10$^{-5}$ & 8.1 10$^{33}$ & - \\ 
\hline 
HD 161103    & B2V-IIIe& 8.4-8.7& 0.44 & $-$0.64 & 0.69        & $-$32      & 0.8 & 4.5 10$^{-6}$ & 1.0 10$^{32}$ & - \\ 
SAO 49725   &B0.5V-IIIe& 9.23   & 0.38 & $-$0.65 & 0.67        & $-$30      & 2.2 & 2.6 10$^{-6}$ & 5.0 10$^{32}$ & - \\ 
\end{tabular}
\end{table*}

\subsection{Comparison with previously known OB/X-ray binaries}

Among the 13 massive X-ray binaries known before the launch of ROSAT and detected by
the SASS analysis in the galactic plane survey, there are 8 Be/X-ray systems and 5
disk-fed or supergiant wind-fed binaries (see Table \ref{knownxrbs}). Most of the
detected pre-ROSAT Be/X-ray systems are transient sources and apart from EXO 2030+375
which was probably caught in weak outburst (Mavromatakis 1994) all other systems have
X-ray luminosities typical of the quiescent state (L$_{\rm X}$ = 5 10$^{33}$ - 5
10$^{34}$\,\ergs \ in the energy range 0.1-2.4\,keV and corrected for photoelectric
absorption).

The distribution in optical spectral types of the 7 new massive X-ray binary candidates
compares well with that of the previously known systems. In particular, we do not find
accreting candidates with spectral types later than B2 and having X-ray luminosities
above 10$^{32}$\,\ergs \ among the candidates earlier than B6, although the possibility
to detect an excess of X-ray luminosity is in principle larger for the later stars. The
absence of good accreting candidates later than B2 is also not an artifact of our
selection criteria since our threshold of \Lx/\Lbol \ = 3 10$^{-6}$ implies a limiting
\Lx \ of $\approx$ 10$^{31}$\,\ergs \ at B5V and 7 10$^{31}$\,\ergs \ at B2V.  In fact we
do detect an excess of X-ray emission in the range of 2-7 10$^{31}$\,\ergs\ from four
B3V-B5V stars (HD 38087, HD 38023, HD 36262 and HD 53339). The youth of the OB
associations in which these four extreme stars are located favours an explanation in
terms of X-ray activity from a pre-main sequence low mass companion.  
 
The distribution in X-ray luminosities of the four best Be/X-ray candidates is also
comparable with that of known Be/X-ray systems in quiescence. On the other hand, the
two additional candidates, HD 161103 and SAO 49725 have significantly lower X-ray
luminosities.

The un-absorbed X-ray luminosities of the five first rank candidates clearly indicate
that the accreting object is a neutron star or less probably a black hole. All five
sources have HR2 $\geq$ 0.5 similar to that exhibited by the majority of the known
X-ray binaries, probably reflecting the hard intrinsic energy distribution. 
Unfortunately, in all cases, at most $\approx$ 100 photons were collected from the new
sources preventing more detailed spectral modelling.  Binary stellar evolution models
predict the existence of five times more Be + white dwarf than Be + neutron star
binaries (Pols et al.  1991).  Waters et al. (1989) compute that white dwarfs
accreting from a Be envelope should display X-ray luminosities in the range from
10$^{29}$ to 10$^{33}$ erg s$^{-1}$. So far there is no established Be + white dwarf
systems. Using ROSAT PSPC observations Haberl (1995) argues that $\gamma$
Cassiopeiae could well be an example of a white dwarf accreting from the dense
circumstellar envelope of a Be star although the hard X-ray properties of the source
are consistent with those of an accreting neutron star in a widely separated orbit
(White et al. 1982; Waters 1989).  In our sample, only HD 161103 and SAO 49725 have
low enough X-ray luminosities to qualify as Be + white dwarf candidates if their X-ray
luminosity excess is confirmed.  It may be noted that these two candidates have large
Balmer emission revealing the presence of a high density envelope.  A white dwarf
could easily accrete enough matter from such a dense circumstellar material in order
to produce the $\approx$ 10$^{32}$\,\ergs \ detected by ROSAT. The X-ray spectrum of
an accreting white dwarf in a Be envelope may not be necessarily as hard as that of a
neutron star (see discussion in Meurs et al. 1992).

Two quite different mechanisms may account for the huge variations in X-ray
luminosities often observed in these binaries. First, the motion of the compact star
along an eccentric orbit with a period of weeks or longer may produce periodic
outbursts when the X-ray source crosses the densest parts of the envelope close to
periastron.  Second, most Be stars are optically variable and this variability has
usually been attributed to dramatic changes in the size and density of the
equatorially condensed circumstellar envelope responsible for the Balmer and infrared
emission. This second mechanism may produce large variations of the X-ray luminosity
whatever is the orbital phase.  

Our sample of four first rank new Be/X-ray systems probably illustrates all these
possible configurations. Although the time interval between the X-ray and optical
observation may introduce some additional scatter, it may be significant that these four
Be stars display \Halpha \ equivalent widths smaller than those of the previously known
sources exhibiting X-ray outbursts (e.g. A1118$-$61, A0535+26, EXO2030+375).  This
suggests that the weakness of the envelope could to some extent explain the low
persistent X-ray luminosities of LS 992 (\Lx \ $\leq$ 1.4 10$^{33}$\,\ergs ) and the absence of
strong recorded outbursts from BSD 24- 491.  On the other hand the relatively weak
\Halpha \ emitting Be star LS 1698 had a recorded bright outburst in the early seventies
indicating that some of our new Be/X-ray sources could be hard X-ray transients in the
quiescent state.  These sources may undergo an outburst on the occasion of a future
ejection of matter in the circumstellar envelope.  LS 992 may be a particular case since
it exhibited a factor 100 variation between survey and pointed observations.  Optical
spectroscopy contemporaneous to the pointed low state observations shows relatively weak
\Halpha \ emission at the same level as 7 months before. We could have here a low peak
X-ray luminosity outbursting system where the main mechanism for variability is orbital
motion. During the follow-up pointed observations two among of our seven new candidate
X-ray binaries (LS 992 and LS 1698) displayed a PSPC count rate below the detection 
threshold of the survey.
This illustrates the strong variability of these sources and the well known fact that
any new X-ray survey leads to the discovery of new members of this class.

\subsection{Hard X-ray emission}

A soft thermal bremsstrahlung component with a luminosity of typically 10$^{34}$ erg
s$^{-1}$ is observed from massive X-ray binaries with a supergiant counterpart (Haberl
et al. 1994). This emission arises probably in a bow shock around the neutron star
traveling through the dense stellar wind of the star. The X-ray luminosity of
LS 5039 observed by ROSAT might be fully accounted for by this thermal emission. Also
the hardness ratios are consistent with the values found for Vela X-1 from a
pointed ROSAT observation (HR1 = 0.99$\pm$0.01 and HR2 = 0.62$\pm$0.01). The relation
between X-ray luminosity in the thermal component and the luminosity in the hard
spectrum originating at the neutron star strongly depends on the wind density and the
system parameters and it is therefore difficult to predict the luminosity at higher
energies.

Be/X-ray binaries on the other hand rarely show a soft component. For instance, the
X-ray spectrum from X Persei is well represented by a power law with exponential
high-energy cutoff between 0.1 and 12\,keV as the combined ROSAT and BBXRT results show
(Haberl 1994). For these systems the spectra might be extrapolated to higher energies,
however the cutoff energy lies in general outside the ROSAT band and is not known.

A mean colour excess of E(B-V) = 1.0 for the new OB/X-ray binaries corresponds to \nh
\ = 5.5 10$^{21}$\,cm$^{-2}$ (Predehl \& Schmitt 1995). Assuming a power law spectrum
with a photon index in the range of 0 - 2 and a cutoff energy $\geq$ 10\,keV (White et
al. 1983, Tanaka 1986) implies that 1 PSPC count\,s$^{-1}$ corresponds to an
absorbed 2-10\,keV flux in the range of 4 10$^{-11}$- 7 10$^{-10}$
erg\,cm$^{-2}$\,s$^{-1}$ or $\approx$ 2-30 Uhuru count s$^{-1}$ (Forman et al. 1978). 
The presence of a soft component will obviously decrease the hard X-ray flux
corresponding to a given PSPC count rate.  Therefore, with an estimated limiting count
rate of $\approx$ 1 Uhuru count s$^{-1}$, highly depending on confusion effects in the
galactic plane region, the new X-ray sources were most probably below the detection
threshold of the all-sky surveys carried out by Uhuru and Ariel V (except for LS 1698
= 4U 1036$-$56 and LS I +61 235 which contributes to the Uhuru source 4U 0142+61). The
absence of detection in the HEAO A-1 X-ray catalogue (Wood et al.  1984) above a
2-10\,keV flux of $\approx$ 5 10 $^{-12}$ erg\,cm$^{-2}$\,s$^{-1}$ is also compatible
with the recorded PSPC count rates.

\subsection{Distribution of the new sources in the Galaxy}

The distribution of the new massive X-ray binary candidates in the Galaxy follows the spiral
arm structures traced by \ion{H}{II}\ regions (Georgelin \& Georgelin 1976) and open
clusters (Vogt \& Moffat 1975).  SAO 49725 and LS 992 are located on the local arm in
opposite directions as seen from the Sun, BSD 24- 491 and LS I +61 235 are in the Perseus
arm and LS 1698, HD 161103 and LS 5039 probably all belong to the Sagittarius-Carina arm. 
Comparing with optical absorption maps from Neckel \& Klare (1980) shows that all new ROSAT
detected massive X-ray binary candidates are in regions of relatively low interstellar
absorption. This is not surprising considering the sensitivity of the PSPC count rate to
interstellar extinction and the selection resulting from the correlation with optical
catalogues. Stars with detailed spectral types are probably complete down to B $\approx$
9-10 and the Luminous Star catalogue out of which about half of our optical input sample was
extracted is complete down to B $\approx$ 12. Therefore, the optical identifications
reported here tend to be close to the completeness level of the optical input catalogue,
clearly suggesting that several other systems with similar X-ray luminosities but without
optical entries, because associated with fainter optical objects, may well be present in the
ROSAT survey. This is consistent with the fact that among the 13 known massive X-ray
binaries listed in Table \ref{knownxrbs}, 5 had no known entries in optical
catalogues before their X-ray discovery.  In a future paper we will report on some of these new ROSAT discoveries.

In spite of the patchiness of interstellar absorption and incompleteness of the
optical catalogues it may be possible to estimate the space density of 
massive X-ray binaries (see e.g. Meurs \& van den Heuvel 1989) and the low end of
their X-ray luminosity function using the ROSAT all-sky survey.  Such a study may 
allow to constrain the contribution of this population to the overall hard X-ray
emission of the Galaxy and more precisely to the hard X-ray galactic ridge detected by
EXOSAT (Warwick et al. 1985).

\begin{table*}
  \caption{Optical and X-ray characteristics of the known OB/X-ray systems 
detected by ROSAT during the all-sky survey at $|b| \ \leq 20$\degree . 
About 5\% of the galactic sky is not covered by this study (see Section 2.1). 
Distances, \Lx/\Lbol \ and \Lx \ are computed using the automatic process 
described in section  2.4 and the results of the SASS analysis. 
The first group consists of Be/X-ray systems and the second group 
gathers disk and wind fed high mass X-ray binaries}
    \label{knownxrbs}
\begin{tabular}{llllllll}
Source Name                  & Spectral & EW(\Halpha) & d    & \Lx/\Lbol     & \Lx          &   HR1      & HR2 \\
                             & Type     & range (\AA) &(kpc) &               & (\ergs )     &            &            \\
\hline
LS 437 = A0726$-$260         & B0-1e    & $-$5.5      & 8.0 & 1.1 10$^{-4}$ & 4.9 10$^{34}$ & 0.76$\pm$0.16 & 0.45$\pm$0.29\\
He 3 -640 = A1118$-$61       & O9.5V-IVe&$-$50 \ $-$60& 3.2 & 1.8 10$^{-5}$ & 5.4 10$^{33}$ & 0.62$\pm$0.43 & 0.59$\pm$0.46\\
1H1909+096                   & Be       & $-$16       & 3.2  & 1.7 10$^{-5}$ & 2.5 10$^{34}$ & 0.87$\pm$0.12 & 0.85$\pm$0.14\\
EXO 2030+375                 & Be       &$-$9 \ $-$15 & 5.6  & 1.1 10$^{-3}$ & 4.9 10$^{35}$ & 0.97$\pm$0.03 & 0.95$\pm$0.04\\
LS I +65 10 = 3A 0114+650    & B0.5IIIe & $-$2.1      & 2.6  & 5.2 10$^{-5}$ & 1.7 10$^{34}$ & 1.00           & 0.82$\pm$0.16\\
V615 Cas = 1E 0236.6+6100    & B0.5V-IIIe&            & 3.1  & 4.4 10$^{-5}$ & 2.0 10$^{34}$ & 0.66$\pm$0.29 & 0.03$\pm$0.71\\
HR 1209 = X Per              & O9.5V-IIIe& 0 \ $-$15    &  0.76  & 1.7 10$^{-5}$ & 9.1 10$^{33}$ & 0.96$\pm$0.04 & 0.39$\pm$0.03\\
HD   245770 = A 0535+26      & O9.7Ve   &$-$15 \ $-$20& 2.7  & 3.8 10$^{-5}$ & 1.7 10$^{34}$ & 0.91$\pm$0.09 & 0.34$\pm$0.10\\
\hline
HD 226868 = Cyg X-1          & O9.7Iab  &             & 2.3  &1.8 10$^{-3}$ & 2.7 10$^{36}$ & 0.99$\pm$0.01 & 0.47$\pm$0.01\\
HD 77581 = Vela X-1          & B0.5Ib   &             & 1.8  &1.8 10$^{-5}$ & 2.7 10$^{34}$ & 1.00           & 0.64$\pm$0.03\\
HD   153919 = 4U 1700$-$37   & O6.5Iaf+ &             & 1.1  &5.4 10$^{-6}$ & 6.0 10$^{33}$ & 0.97$\pm$0.03 & 0.91$\pm$0.03\\
Cen X-3                      & O6.5II-III&            &13.0  &1.0 10$^{-3}$ & 1.1 10$^{36}$ & 0.91$\pm$0.09 & 0.60$\pm$0.11\\
V830 Cen = 1E1145.1$-$6141   & B2Iae    &             & 4.5  &2.2 10$^{-4}$ & 9.1 10$^{34}$ & 0.92$\pm$0.08 & 0.80$\pm$0.18\\
\end{tabular}
\end{table*}

\section{Conclusions}

Using the cross-correlation in position of the OB star catalogues held in SIMBAD with
the low galactic latitude part of the RASS we selected 24 early type stars which
apparently exhibited a soft X-ray excess over the normal stellar level. Follow-up
optical and X-ray observations allowed to strongly suggest the presence of a compact
object, probably a neutron star, in four cases and establish the X-ray binary nature 
of LS I +61 235.  In two additional cases which 
still require confirmation the X-ray luminosity may be
compatible with that expected from an accreting white dwarf.  

The new massive X-ray binary candidates are clearly representative of the low end of
the X-ray luminosity function. Their soft X-ray luminosities are comparable to those
of hard X-ray transients in quiescence and those of the low luminosity Be/X-ray
systems such as X Persei. One of the new Be/X-ray systems is the probable counterpart
of 4U 1036$-$56 which was in outburst during the years 1970-1976.

Four B stars located in the Orion and Canis Major OB associations exhibit X-ray
luminosities in the range of 2-7 10$^{31}$\,\ergs \ well above the expected normal
stellar emission. Such X-ray luminosities appear at the high end of the luminosity
distribution of active stars and the usual interpretation in terms of a young late
type companion star may not hold. 

Finally we report the probable discovery of a white dwarf companion to the B5V type
star HR 2875. 

\begin{acknowledgements} We thank N. Fourniol, P. Guillout and the night assistants at
Observatoire de Haute-Provence for carrying out some of the observations at the 1.2\,m and
1.93\,m telescope. We also thank S.A. Gol\c{c}alves and R.D.A. da Costa for carrying out
some of the observations at the 1.6\,m of L.N.A.  The ROSAT project is supported by the
German Bundesministerium f\"ur Bildung, Wissenschaft, Forschung und Technologie (BMBF/DARA)
and the Max-Planck-Gesellschaft.  C.M. acknowledges support from a CNRS-MPG cooperation
contract and thanks Prof. J.  Tr\"umper and the ROSAT group for their hospitality and
fruitful discussions.  We acknowledge detailed and useful comments from an anonymous
referee.  GSC data were extracted from the STARCAT facility at ESO (Pirenne et al. 1993)
and later from the SIMBAD catalogue GSC browser (Preite-Martinez \& Ochsenbein 1993). This
research has made large use of the SIMBAD database operated at CDS, Strasbourg, France.
 
\end{acknowledgements}

\end{document}